%% file: aaai.tex
\documentclass[letterpaper]{article} % DO NOT CHANGE THIS
\usepackage{aaai2026}
\usepackage{times}  % DO NOT CHANGE THIS
\usepackage{helvet}  % DO NOT CHANGE THIS
\usepackage{courier}  % DO NOT CHANGE THIS
\usepackage[hyphens]{url}  % DO NOT CHANGE THIS
\usepackage{graphicx} % DO NOT CHANGE THIS
\usepackage{natbib}  % DO NOT CHANGE THIS AND DO NOT ADD ANY OPTIONS TO IT
\usepackage{caption} % DO NOT CHANGE THIS AND DO NOT ADD ANY OPTIONS TO IT
\usepackage{amssymb}
\usepackage{subcaption}
\newcommand{\shortname}{\textit{RecCocktail}}
\usepackage{paralist}
\usepackage{graphicx}
\usepackage{enumitem}
\usepackage{multirow}
\usepackage{algorithm}
\usepackage{algorithmic}
\usepackage{adjustbox}
\usepackage{booktabs}
\usepackage[table]{xcolor}
\usepackage{amsmath}
\usepackage{newfloat}
\usepackage{listings}
\DeclareCaptionStyle{ruled}{labelfont=normalfont,labelsep=colon,strut=off} % DO NOT CHANGE THIS
\floatstyle{ruled}
\newfloat{listing}{tb}{lst}{}
\floatname{listing}{Listing}

\title{\textit{RecCocktail}: A Generalizable and Efficient Framework for \\LLM-Based Recommendation}

\author{
    Min Hou\textsuperscript{\rm 1}, Chenxi Bai\textsuperscript{\rm 1}, Le Wu\textsuperscript{\rm 1}\thanks{Corresponding Author.}, Hao Liu\textsuperscript{\rm 1}, Kai Zhang\textsuperscript{\rm 2},\\ Weiwen Liu\textsuperscript{\rm 3}, Richang Hong\textsuperscript{\rm 1}, Ruiming Tang\textsuperscript{\rm 4}, Meng Wang\textsuperscript{\rm 1}
}
\affiliations{
    \textsuperscript{\rm 1}Hefei University of Technology,
    \textsuperscript{\rm 2}University of Science and Technology of China,\\
    \textsuperscript{\rm 3}Shanghai Jiao Tong University,
    \textsuperscript{\rm 4}Kuaishou Inc.
    	hmhoumin@gmail.com, 
        bcx@mail.hfut.edu.cn, 	lewu.ustc@gmail.com,
        hliu7879@163.com,
        kkzhang08@ustc.edu.cn,
        wwliu@sjtu.edu.cn,
        hongrc.hfut@gmail.com,
        tangruiming@kuaishou.com,
        eric.mengwang@gmail.com

}

\usepackage{bibentry}
\begin{document}

\maketitle

\begin{abstract}
Large Language Models (LLMs) have achieved remarkable success in recent years, owing to their impressive generalization capabilities and rich world knowledge. To capitalize on the potential of using LLMs as recommender systems, mainstream approaches typically focus on two paradigms. The first paradigm designs multi-domain or multi-task instruction data for generalizable recommendation, so as to align LLMs with general recommendation areas and deal with cold-start recommendation. The second paradigm focuses on enhancing domain-specific recommendation tasks, improving performance in warm recommendation scenarios.  While most previous works treat these two paradigms separately, we argue that they have complementary advantages, and combining them can yield better results.
In this paper, we propose a generalizable and efficient LLM-based recommendation framework \shortname. Our approach begins with fine-tuning a ``base spirit" LoRA module using domain-general recommendation instruction data to align LLM with recommendation knowledge. Next, given users' behavior of a specific domain, we construct a domain-specific ``ingredient" LoRA module. We then provide an entropy-guided adaptive merging method to mix the ``base spirit" and the ``ingredient" in the weight space. Please note that, \shortname~ combines the advantages of the existing two paradigms without introducing additional time or space overhead during the inference phase. Moreover,
\shortname~is efficient with plug and play, as the ``base spirit" LoRA is trained only once, and any domain-specific ``ingredient" can be efficiently mixed with only domain-specific fine-tuning. Extensive experiments on multiple datasets under both warm and cold-start recommendation scenarios validate the effectiveness and generality of the proposed \shortname. 
%Codes are available at \url{https://anonymous.4open.science/r/RecCocktail}. 
%To the best of our knowledge, this work is the first attempt that incorporates the model merging into the recommender system tasks. It solves the problem of conflict between generalization and efficiency in large model recommendation fields.
\end{abstract}
\begin{links}
    \link{Code}{https://anonymous.4open.science/r/RecCocktail}
    % \link{Datasets}{https://aaai.org/example/datasets}
    % \link{Extended version}{https://aaai.org/example/extended-version}
\end{links}
\input{sections/intro}

\input{sections/preliminary}

\input{sections/method}
\input{sections/experiments}

\input{sections/conclusion}

\bibliography{aaai2026}

\appendix
\input{sections/appendix}

\end{document}

%% file: sections/intro.tex
\section{Introduction}
Large Language Models~(LLMs) have demonstrated significant success across diverse fields~\cite{zhao2023survey}, driven by their emergent capabilities~\cite{dong2024survey,huang-chang-2023-towards} such as world knowledge, language understanding, and complex reasoning. 
Recently, LLMs have introduced transformative advancements to recommendation tasks~\cite{10.1145/3640457.3688104,deng2022toward,geng2022recommendation}. 
\iffalse
Notably, LLMs have shown potential in capturing nuanced item semantics~\cite{10.1145/3640457.3688104}, understanding diverse user interests~\cite{deng2022toward}, and unifying various recommendation tasks~\cite{geng2022recommendation}.
These advancements highlight the promise of utilizing LLMs as recommender systems, positioning LLM-based recommendations as a compelling area for further exploration.
\fi
\begin{figure*}[ht]
    \centering \includegraphics[width=0.98\textwidth]{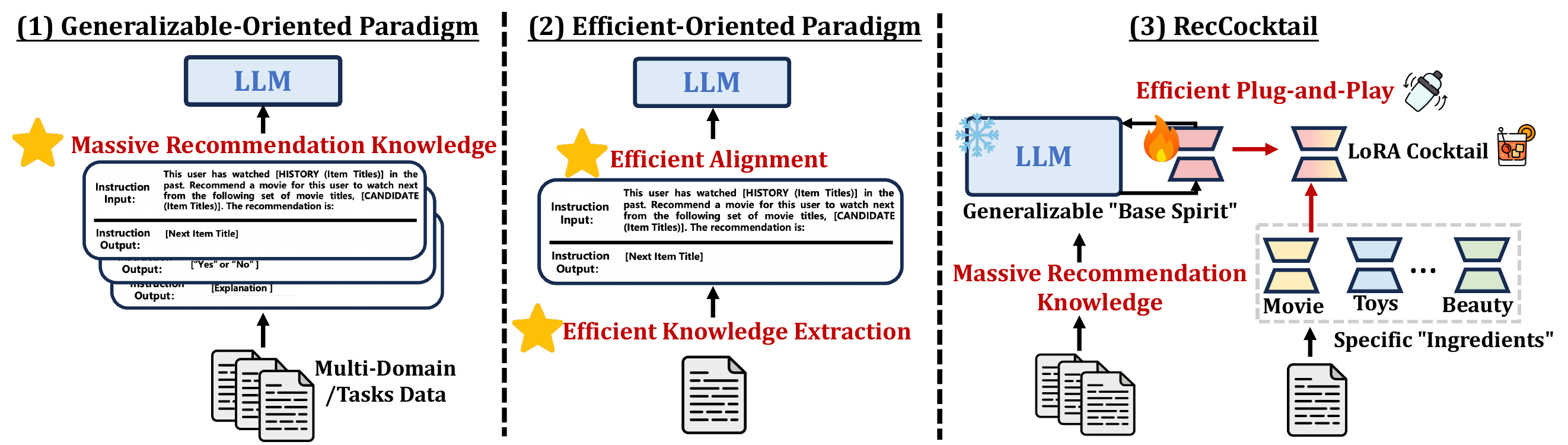}
    %\vspace{-0.4cm}
    \caption{Illustration of different LLM-based recommendation paradigms. (1) Breadth-oriented paradigm. (2) Depth-oriented paradigm. (3) Our proposed \shortname.}
    \label{fig:intro}
\end{figure*}
Along this line, the emergence of ChatGPT and its remarkable reasoning capabilities have catalyzed early studies~\cite{dai2023uncovering,sanner2023large,wang-etal-2023-rethinking-evaluation}. These works focus on the zero-shot/few-shot recommendation potential of LLMs through in-context learning~\cite{dong2024survey}. However, the intrinsic gap between the pre-training general text corpus of LLMs and the requirements of recommendation tasks results in suboptimal performance when relying solely on in-context learning. Consequently, the key to developing an effective LLM-based recommender system lies in bridging this gap, enabling the model to truly ``understand" how to recommend.

To address this challenge, researchers have proposed a variety of approaches. We classify them into two paradigms, each tackling the problem from a distinct perspective.
As shown in Figure \ref{fig:intro}(1), the first one is summarized as the \textbf{breadth-oriented paradigm}. These works integrate multi-domain~\cite{10.1145/3705727} or multi-task~\cite{geng2022recommendation,10.1145/3708882,cui2022m6} recommendation data to construct extensive recommendation world knowledge, paving the way for developing a generalizable LLM-based recommender. 
The key focus of this paradigm is the integration of multi-source data to build instruction-tuning datasets~\cite{peng2024ecellm,jin2023amazonm} and the design of instruction templates~\cite{10.1145/3708882,geng2022recommendation} tailored to various tasks.
%The research focus of paradigm 1 lies in integrating multi-source data to construct instruction tuning datasets~\cite{peng2024ecellm,jin2023amazonm} and designing instruction templates~\cite{10.1145/3708882,geng2022recommendation} specifically tailored to different tasks. 
The second paradigm is termed the \textbf{depth-oriented paradigm}, illustrated in Figure \ref{fig:intro}(2). This line of research seeks to enable LLMs to deeply comprehend recommendation tasks within specific domains. Key areas of focus include: the in-depth extraction of domain-specific recommendation knowledge, such as collaborative filtering information~\cite{lin2024bridging,10.1145,10.1145/3626772.3657690,10.1145/3589334.3645458,kong2024customizing}, and the development of efficient and effective alignment methods between LLMs and recommendation tasks. Specifically, compared with enormous parameters in LLMs, downstream tasks do not have sufficient data for tuning all parameters. 
Therefore, parameter-efficient fine-tuning methods become optimal for applying LLMs, in which lightweight Low-Rank Adapter (LoRA) is one representative work~\cite{hu2022lora}. By borrowing ideas of LLMs, these methods include leveraging~\cite{bao2023tallrec} or enhancing~\cite{kong2024customizing} LoRA fine-tuning techniques and designing data-efficient fine-tuning strategies~\cite{10.1145/3626772.3657807}.

These works make significant advancements in recommendation research. Nevertheless, we argue that these two paradigms have complementary advantages. 
\iffalse
The first paradigm masters general recommendation knowledge and can be well generalized to various recommendation scenarios. The second paradigm learns each user's unique preference and is suitable for the warm-start recommendation of a specific domain. However, the specific recommendation domain performance could not be easily transferred to other domains. In fact,
\fi
Both generalizable recommendation knowledge and efficient domain-specific understanding are essential for recommender systems. 
Relying solely on one aspect risks falling short in addressing the diverse challenges in real-world recommendation scenarios.
The breadth-oriented paradigm may underperform in specific domains. Conversely, the depth-oriented paradigm struggles with distribution shifts between training and test data. It faces challenges when new users or items appear or when training data are sparse.

To this end, we investigate how to integrate the advantages of both paradigms to simultaneously enhance the model's generalization ability and domain-specific performance. The task creates significant obstacles: (1) Efficiency. Integrating two paradigms may introduce model complexity, and finding an efficient integrating method without excessive computational overhead is a critical challenge. (2) Generalizability. We need to preserve the model's generalization ability to a large extent, enabling it to quickly scale to new domains, new items, and other new recommendation scenarios.
In this paper, we propose a generalizable and efficient recommendation framework named \shortname, inspired by the cocktail preparation process, as shown in Figure \ref{fig:intro}(3).
Specifically, to align LLM with any recommendation task, \shortname~constructs a general recommendation instruction dataset from multiple recommendation domains, and fine-tunes LLM to get a domain-general LoRA module as ``base spirit". Secondly, to tailor the framework for specific domains, \shortname~constructs domain-specific instruction datasets, and fine-tunes LLM to get a domain-specific LoRA module as ``ingredient". After that, \shortname~performs a highly efficient and effective linear arithmetic operation to merge the ``base spirit" and the ``ingredient" LoRA within the weight space, allowing \shortname~to maintain strong recommendation performance across both specific domains and out-of-distribution scenarios. To further enhance the merging process, we also introduce an adaptive merging method guided by entropy minimization during test time.  Importantly, \shortname~does not introduce additional time or space overhead during the inference phase. Furthermore, \shortname~is designed for ease of use, allowing for a plug-and-play integration where the ``base spirit" is trained once, and domain-specific ``ingredients" are incorporated through minimal fine-tuning. Finally, extensive experiments conducted on various datasets demonstrate the effectiveness and generalizability of the framework in multiple recommendation scenarios, highlighting its potential for broad application.

%% file: sections/preliminary.tex
\section{Preliminary}
\iffalse
In this section, we introduce key concepts underpinning our methodology. First, we cover the task formulation and instruction tuning for LLM-based recommender models. Next, we highlight the use of LoRA for parameter-efficient fine-tuning of LLMs.
\fi
\subsection{LLM-Based Recommendation}
$\bullet$ \textbf{Task Formulation.}
%\textbf{Task Formulation.}
\iffalse
We mainly focus on the sequential recommendation task, which holds significant practical importance. 
\fi
Let $\mathcal{U}$ and $\mathcal{I}$ represent the sets of users and items, respectively. The historical interaction sequence of a user $u \in \mathcal{U}$ is denoted as $\mathcal{S}_u=\left[i_u^1, i_u^2, \ldots, i_u^L\right]$, arranged in chronological order, where $i_u \in \mathcal{I}$ and $L=|\mathcal{S}_u|$. The goal is to predict this user's next liked item $i_u^{L+1} \in \mathcal{I}$ based on the historical interactions.

%\noindent$\bullet$ \quad\textbf{Instruction Tuning for LLM-Based Recommendation.}
\noindent$\bullet$ \textbf{Instruction Tuning for LLM-Based Recommendation.}
For LLM-based recommendation, instruction tuning~\cite{wei2022finetuned} is the key step to bridge the gap between the next-word prediction objective of LLMs and the recommendation task~\cite{bao2023tallrec,kong2024customizing,10.1145/3626772.3657807}.
Formally, instruction tuning involves fine-tuning LLMs using training data organized as explicit instruction pairs $\{(\mathbf{x}_u, \mathbf{y}_u)|u \in \mathcal{U}\}$. Here, $\mathbf{x}_u$ represents a detailed textual instruction that encapsulates the interaction sequences $\mathcal{S}_u$ and the recommendation task, while $\mathbf{y}_u$ corresponds to the textual description of the predicted item $i_u^{L+1}$. The instruction fine-tuning process is guided by minimizing the following autoregressive loss function:
\begin{equation}
\mathcal{L}_\Theta^{L L M}=-\sum_{u} \sum_{t=1}^{|\mathbf{y}_u|} \log P_\Theta\left(y_u^t \mid \mathbf{y}^{<t}_u, \mathbf{x}_u\right),
\label{eq:ft}
\end{equation}
where $y^t_u$ denotes the $t$-th token of the output sequence $\mathbf{y}_u$, $\mathbf{y}^{<t}_u$ is the token sequence preceding $y^t_u$, and $\Theta$ is the LLM's model parameters.

\subsection{Instruction Tuning with LoRA}
In traditional fine-tuning as described in Eqn. (\ref{eq:ft}), updating all parameters makes the process highly computationally intensive, particularly for LLMs.
To address this issue, parameter-efficient methods are designed to fine-tune LLMs while updating only a small subset of parameters.
Low-Rank Adaptation (LoRA)~\cite{hu2022lora} is the mainstream approach. LoRA addresses this issue by introducing low-rank matrices that are trained alongside the frozen original model weights. This allows the model to adapt to specific tasks by learning a small number of additional parameters, without requiring modifications to the entire model.

Specifically, for any pre-trained weight matrics $\boldsymbol{W}_0 \in \mathbb{R}^{d \times k}$ in the transformer block of the LLM, which takes an input $\boldsymbol{x} \in \mathbb{R}^k$ and output $\boldsymbol{h}$. LoRA modifies  $\boldsymbol{h} = \boldsymbol{W}_0 \boldsymbol{x}$ to:
\begin{equation}
    \boldsymbol{h} = \boldsymbol{W}_0 \boldsymbol{x} + \boldsymbol{B}\boldsymbol{A}\boldsymbol{x},
    \label{eq:lora}
\end{equation}
where $\boldsymbol{B}\in \mathbb{R}^{d \times r}$, $\boldsymbol{A}\in \mathbb{R}^{r \times k}$ are the low-rank projection matrices. Notably, the rank $r \ll \min (d, k)$, ensuring that the number of parameters introduced by $\boldsymbol{B}\boldsymbol{A}$ is significantly fewer than those of $\boldsymbol{W}_0$, as $dr + rk \ll dk$. During fine-tuning, only $\boldsymbol{A}$ and $\boldsymbol{B}$ are updated, while $\boldsymbol{W}_0$ remains fixed. In a similar way, LoRA adapter is generally applicable to any LLM layer desired for updating. The training objective of LoRA fine-tuning can be formulated as:
\begin{equation}
\max _{\Delta\Theta} \sum_{u} \sum_{t=1}^{|\mathbf{y}_u|} \log P_{\Theta_{\text{pre}} +\Delta\Theta }\left(\mathbf{y}_u^t \mid \mathbf{y}^{<t}_u, \mathbf{x}_u\right).
\end{equation}
Here, $\Theta_{\text{pre}}$ is the parameters of the pre-trained LLM. $\Delta\Theta=\{\boldsymbol{A}^l,\boldsymbol{B}^l\}_{l=1}^L$ denotes the set of parameters of LoRA fine-tuning, and $L$ represents the number of LoRA modules.

%% file: sections/method.tex
\section{Methodology}
\label{sec:method}
\iffalse
\begin{figure}[b]
    \centering
     \includegraphics[width=0.48\textwidth]{figures/phase1_3.pdf}
    %\vspace{-0.4cm}
    \caption{Illustration of aligning LLM with recommendation task.}
    \label{fig:phase1}
    \Description{}
\end{figure}
\fi
\begin{figure*}[ht]
    \centering \includegraphics[width=0.99\textwidth]{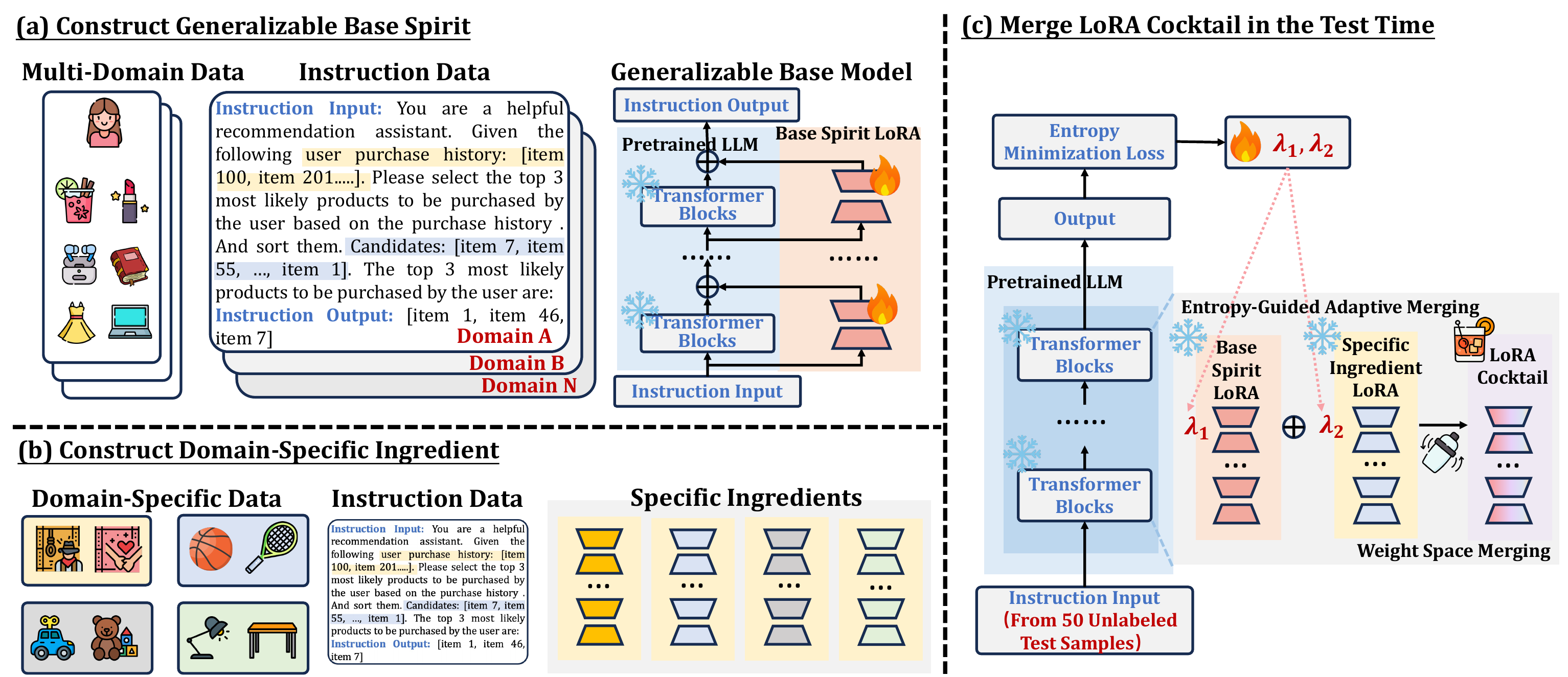}
    \caption{Illustration of our proposed \shortname~framework.}
    \label{fig:framework}
\end{figure*}
In this section, we propose \shortname, a generalizable, effective, and efficient LLM-based recommendation framework. 
As shown in Figure \ref{fig:framework}, \shortname~operates through three key stages. 
We start by preparing the base spirit.
We align the LLM with any recommendation task and fine-tune a domain-general LoRA module as the base spirit~(Section \ref{sec:stage1}). Secondly, to adapt the framework to a specific domain, we fine-tune the domain-specific LoRA module (Section \ref{sec:stage2}) as an ingredient. Subsequently, \shortname~performs an efficient linear arithmetic operation to merge the base spirit LoRA and the ingredient LoRA into a cocktail within the weight space (Section \ref{sec:stage3_1}).
We further provide a test-time entropy-guided adaptive merging method to quickly construct cocktail LoRA tailored to different inference scenarios.
The merged cocktail LoRA collaborative enhances performance across both specific domains and out-of-distribution scenarios without introducing additional time or space overhead.

\iffalse
Notably, our framework is designed for ease of use, offering plug-and-play integration, where the domain-general module is trained once, and domain-specific adaptations are incorporated with minimal fine-tuning. \shortname~ is capable of significantly enhancing the recommendation performance and generalization across various domains at a relatively low cost.
\fi

\iffalse
First, the recommendation ability alignment stage aligns the pre-trained LLM with the general characteristics of recommendation tasks using an autoregressive objective (Section \ref{sec:stage1}). Next, the domain-specific fine-tuning stage incorporates specialized knowledge from specific domains (e.g., movies, beauty products) into the pre-trained LLM (Section \ref{sec:stage2}). Finally, we introduce the Mixture-of-LoRA method to address specific recommendation scenarios by adaptively integrating general and domain-specific knowledge. This approach significantly enhances the model's recommendation performance and generalization ability without increasing inference time or the number of model parameters (Section \ref{sec:stage3_1}).
\fi

%\subsection{Aligning LLM with Recommendation Task}
\subsection{Constructing Generalizable Base Spirit}
\label{sec:stage1}
In this subsection, we construct a generalizable base spirit equipped with general recommendation knowledge.
As illustrated in Figure \ref{fig:framework}(a), we first create a large-scale instruction dataset by aggregating user behavior data from multiple domains.
We then fine-tune a pre-trained LLM on this dataset using the LoRA technique, resulting in a domain-general LoRA module, which serves as our base spirit.

\noindent$\bullet$ \textbf{Instruction Dataset Construction.}
Given $N$ recommendation domains (i.e., $\mathcal{D}^1$, $\mathcal{D}^2$,..., $\mathcal{D}^N$), let $\mathcal{U}^n$, $\mathcal{I}^n$, and $\mathcal{S}^n$ denote the user set, item set, and user interaction sequence set of domain $n$, respectively.
To provide general user modeling and recommendation knowledge, we aggregate data from all $N$ domains and design instruction templates to convert them into text format. Note that both the choice of recommendation domains and the design of instruction templates are flexible. In this paper, we adopt the template shown in Figure \ref{fig:framework}(a).
We transform the multi-domain recommendation data into an instruction dataset $\mathcal{D}_{g}=\{(\textbf{x}, \textbf{y})\}$, where $\textbf{x}$ and $\textbf{y}$ represent the instruction input and output, respectively. The input a task description, the user’s historical interactions, and a set of candidate items, all expressed in text. Each item is represented by its title. The candidate set includes one ground-truth item and several randomly sampled negative samples. The output is designed to rank the user's next most likely item.

\iffalse
We denote the mixed user, item, and the interaction set as $\mathcal{U}$, $\mathcal{I}$, $\mathcal{S}$, respectively.
For each user interaction sequence $\mathcal{S}_u=\left[i_u^1, i_u^2, \ldots, i_u^L\right]$ 

For the instruction template,

To provide general user modeling and recommendation knowledge, we mix user behaviors $\mathcal{S}_u=\left[i_u^1, i_u^2, \ldots, i_u^L\right]$ from multiple domains and reconstruct a recommendation instruction dataset $\mathcal{D}_{g}=\{(\textbf{x},\textbf{y})\}$, where $\textbf{x}$ and $\textbf{y}$ denote the instruction input and output, respectively. The choice of the instruction format and the original user behaviors data can be arbitrary. In this work, we collect 8 domains' user recommendation data from Amazon Review dataset\footnote{\url{https://cseweb.ucsd.edu/~jmcauley/datasets.html}}, including Clothing, Cell, Grocery, Health, Home, Pet, Tools, and Videos. Specifically, as shown in the left part of Figure \ref{fig:phase1}, the instruction input starts from a system prompt that defines the LLM's role-"\textit{You are a helpful recommendation assistant}"
\fi

\noindent$\bullet$ \textbf{Tuning Domain-General LoRA Module.}
Given the instruction dataset $\mathcal{D}_{g}$, we apply LoRA fine-tuning to adapt the pre-trained LLM for general recommendation tasks. The pre-trained model parameters are kept frozen, while trainable low-rank decomposition matrices are introduced into each layer of the Transformer architecture, enabling efficient and lightweight tuning. Formally,
\begin{equation}
\max _{\Delta\Theta_g} \sum_{(\mathbf{x},\mathbf{y}) \in \mathcal{D}_{g}} \sum_{t=1}^{|\mathbf{y}|} \log P_{\Theta_{\text{pre}} +\Delta\Theta_g}\left(\mathbf{y}^t \mid \mathbf{y}^{<t}, \mathbf{x}\right),
\end{equation}
where $\Theta_{\text{pre}}$ is the parameters of the pre-trained LLM, and $\Delta\Theta_g$ denotes the set of parameters of LoRA fine-tuning. By undergoing this fine-tuning step, $\Delta\Theta_g$ is now enriched with extensive general recommendation knowledge.

\subsection{Constructing Domain-Specific Ingredient}
\label{sec:stage2}

Each recommendation domain exhibits unique user behavior patterns, making the acquisition of domain-specific knowledge essential for delivering accurate recommendations. 
%For instance, in the clothing domain, user behavior is predominantly driven by style, whereas in the electronics domain, users are more likely to prioritize compatibility with their existing devices. 
To address these domain-specific characteristics, we construct an instruction dataset $\mathcal{D}_s$ tailored to the target domain $s$, and fine-tune the pre-trained LLM using the LoRA technique.
The domain-specific LoRA module serves as a ingredient of a cocktail.
As shown in Figure \ref{fig:framework}(b), the instruction template and LoRA fine-tuning procedure follow a similar approach to that described in Section \ref{sec:stage1}. Formally, the training objective for the domain-specific LoRA $\Delta\Theta_s$ is defined as:
\begin{equation}
\max _{\Delta\Theta_s} \sum_{(\mathbf{x},\mathbf{y})\in \mathcal{D}_s} \sum_{t=1}^{|\mathbf{y}|} \log P_{\Theta_{\text{pre}} +\Delta\Theta_s}\left(\mathbf{y}^t \mid \mathbf{y}^{<t}, \mathbf{x}\right).
\end{equation}

\subsection{LoRA Cocktail for Plug-and-Play}
\label{sec:stage3_1}
% \begin{wrapfigure}{r}{0.5\textwidth}
%     \centering 
%     \vspace{-1.1em}
%     \includegraphics[width=0.5\textwidth]{figures/task_arithmetic_5.pdf}
%     \caption{Illustration of task arithmetic~\cite{ilharco2023editing}. (a) A task vector is obtained by subtracting the weights of a pre-trained model from the weights of the same model after fine-tuning. (b) Adding task vectors together improves the performance of the pre-trained model on the tasks under consideration.}
%     \label{fig:task_arithmetic}
% \end{wrapfigure}
\begin{figure}[t]
    \centering
    \includegraphics[width=1.0\columnwidth]{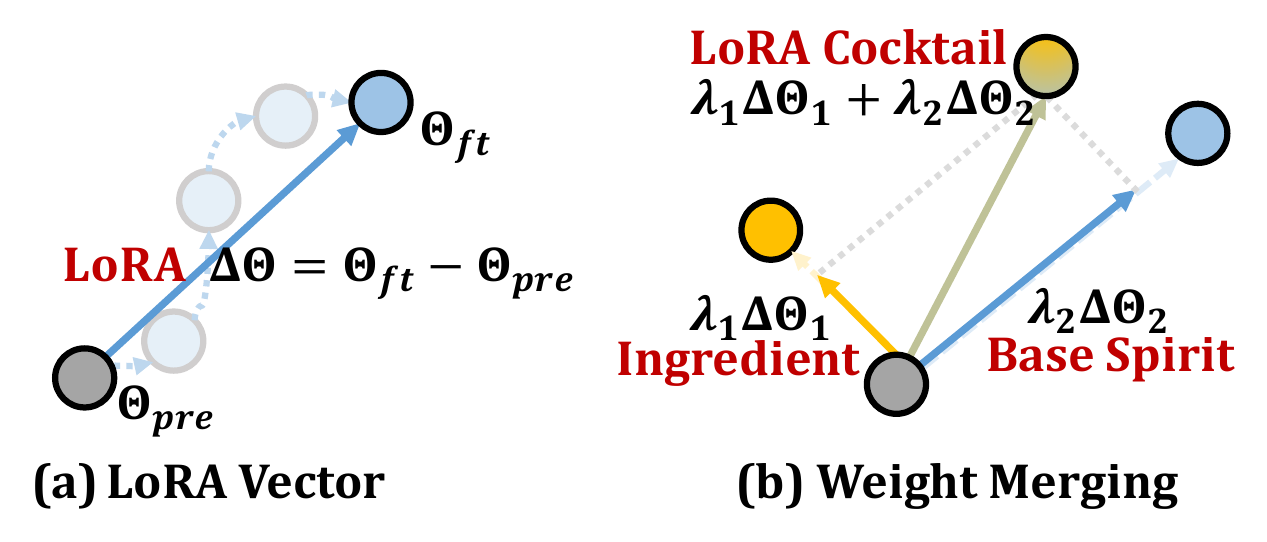}
    \caption{Illustration of task arithmetic~\cite{ilharco2023editing}. (a) Task vectors are obtained by subtracting pre-trained weights from fine-tuned weights. (b) Adding task vectors improves performance on multiple tasks.}
    \label{fig:task_arithmetic}
\end{figure}

After preparing the base spirit and the ingredient, we propose integrating these two parts to improve recommendation accuracy and enhance generalization capabilities simultaneously.
Natural questions arise: could this goal be achieved by applying traditional ensemble learning methods that combine the outputs of multiple models? Or could we integrate general and domain-specific knowledge by directly performing second-round fine-tuning on the domain-general LoRA module with the domain-specific dataset?
Unfortunately, the answer is no. Since LLMs generate natural language text, ensembling their outputs can introduce semantic inconsistencies or ambiguities, while also increasing inference time and GPU memory usage. Meanwhile, performing a second round of fine-tuning risks catastrophic forgetting~\cite{kirkpatrick2017overcoming}, causing the model to collapse.

\noindent$\bullet$ \textbf{LoRA Cocktail.} We propose a simple yet effective method called LoRA cocktail, which linearly merges the model parameters of the domain-general base spirit LoRA $\Delta\Theta_g$ and the domain-specific ingredient LoRA $\Delta\Theta_s$.
We draw an analogy to cocktail making, where mixing a base spirit with an additional ingredient results in a drink that combines the flavors of both parts.
Formally, given the base spirit $\Delta\Theta_g =\{\boldsymbol{A}_g^l,\boldsymbol{B}_g^l\}_{l=1}^L$ and a ingredient $\Delta\Theta_s=\{\boldsymbol{A}_s^l,\boldsymbol{B}_s^l\}_{l=1}^L$, we define the LoRA cocktail operator $\oplus$ as:
\begin{equation}
    \Delta\Theta_m = (\lambda_1\Delta\Theta_g) \oplus (\lambda_2\Delta\Theta_s) = \{\boldsymbol{A}_m^l, \boldsymbol{B}_m^l\}_{l=1}^L,
\label{eq:merge}
\end{equation}
\begin{equation}
    \boldsymbol{A}_m^l = \lambda_1\boldsymbol{A}_g^l + \lambda_2\boldsymbol{A}_s^l, \quad
    \boldsymbol{B}_m^l = \lambda_1\boldsymbol{B}_g^l + \lambda_2\boldsymbol{B}_s^l,
\end{equation}
\iffalse
\begin{equation}
    \boldsymbol{B}_m^l = \lambda_1\boldsymbol{B}_g^l + \lambda_2\boldsymbol{B}_s^l,
\end{equation}
\fi
where the coefficients $\lambda_1$ and $\lambda_2$ represents the importance of merging. We constraint $\lambda_1 + \lambda_2=1$ and $0 <= \lambda_1,\lambda_2 <=1$. They can be considered hyperparameters and selected using the validation data. Please note that the mixed $\Delta\Theta_m$ maintains the same total number of parameters as one standard LoRA, making our LoRA Cocktail method simple, fast, and effective. There is no extra cost at inference time in terms of memory or compute, since we only do element-wise operations on model weights. In addition, the domain-general LoRA module is reusable. When facing a new domain, it is only necessary to retrain a domain-specific LoRA.

LoRA cocktail is the weight merging of the domain-general LoRA module and the domain-specific one. Here we explain how it works as shown in Figure \ref{fig:task_arithmetic}. 
\iffalse
According to Eqn. (\ref{eq:lora}) and (\ref), $\boldsymbol{h} = \boldsymbol{W}_0 \boldsymbol{x} + \boldsymbol{B}\boldsymbol{A}\boldsymbol{x}$,LoRA technique
\fi
We find that after fine-tuning, a LoRA module can be interpreted as a task-specific weight update vector: $\Delta\Theta = \Theta_{\text{ft}} - \Theta_{\text{pre}}$, which defines a direction in the pre-trained model’s weight space. Shifting the model’s weights along this direction enhances its performance on the corresponding task.
Intuitively, a LoRA vector encodes all of the information needed to solve a task that is learned via fine-tuning.
Adding two directions allows the model to move in a way that incorporates both general knowledge and domain-specific knowledge.
This idea aligns with the concept of ``task vectors" proposed in recent work~\cite{ilharco2023editing}, where weight differences from pre-trained to fine-tuned models are shown to encode task-specific information. These vectors, residing in the same parameter space, can be combined to equip the model with multiple capabilities. This is further supported by findings that models fine-tuned from the same initialization often lie in the same error basin~\cite{neyshabur2020being,zhang2023composing}, making such linear merging feasible and effective.
In this light, the LoRA cocktail approach leverages LoRA as task vectors, enabling efficient knowledge integration through simple vector addition.

\iffalse
is inspired by recent studies~\cite{ilharco2023editing,wortsman2022model} on the linear connectivity of trained models in a full finetuning setting. These studies suggest that parameters of tuned models can be directly added to improve generalization, provided they are initialized from the same pre-trained model checkpoint. 
Specifically, as shown in Figure , recent research~\cite{ilharco2023editing} defines the concept of "task vector". A task vector $\Delta\Theta$ specifies a direction in the weight space of a pre-trained model, such that movement in that direction improves performance on the task. It is built by subtracting the weights of a pre-trained model ($\Delta\Theta_{pre}$) from the weights of the same model after fine-tuning ($\Delta\Theta_{ft}$) on a task. Adding task vectors together can improve performance on multiple tasks at once.
The underlying hypothesis is that two models finetuned from the same pre-trained checkpoint often lie in the same error basin~\cite{neyshabur2020being,zhang2023composing}, and thus the parameters could be directly added.
Extending this property to the context of LoRA, we hypothesize that LoRA modules can also be linearly combined. This is because a LoRA module can be considered as the difference between a fine-tuned LLM and its pre-trained counterpart, making it analogous to a task vector.
\fi

\noindent$\bullet$ \textbf{Entropy-Guided Adaptive Merging.}
As shown in Figure \ref{fig:framework}(c), we provide an efficient and automatic way to better choose merging coefficients $\lambda_1$ and $\lambda_2$. As discussed in the previous subsection, $\lambda_1$ and $\lambda_2$ can be chosen by employing the grid-search in the validation data. Nevertheless, (1) it is still lacking a guiding principle. (2) When the distribution of the inference data differs significantly from that of the validation set, the chosen coefficients may perform poorly.
To this end, we introduce entropy minimization on the unlabeled test samples as an optimization surrogate objective to update $\lambda_1$ and $\lambda_2$.
Specifically, the Shannon Entropy~\cite{shannon1948mathematical} is a well-known measure of uncertainty. For a sample $\mathbf{x}_i$, the predicted output of a neural network $\mathcal{F}_\theta(\mathbf{x}_i)$ is $\hat{\mathbf{y}}_i$, the Shannon entropy is calculated as $H(\hat{\mathbf{y}}_i) = -\sum_c^C p\left(\hat{\mathbf{y}}_{i, c}\right) \log p\left(\hat{\mathbf{y}}_{i, c}\right)$, where $p\left(\hat{\mathbf{y}}_{i, c}\right)$ denotes the probability that the input $\mathbf{x}_i$ is predicted to be the $c$-th class. 
Lower entropy indicates that the model has lower uncertainty about its predictions, meaning the model is more confident in its outputs. 
Therefore, the intuition behind our method is that the good coefficients $\lambda_1$ and $\lambda_2$ for the test inputs should make the mixed model more confident in its prediction, that is, it should lead to lower model entropy over the input~\cite{wang2021tent,wang2021emea,yang2024adamerging}.
Formally, we collect a set of unlabeled test samples $\mathcal{D}_t$, i.e., some instruction inputs in the test time. We fix the $\Delta\Theta_g$, $\Delta\Theta_s$, $\Theta_\text{pre}$, and using the following entropy minimization loss to update coefficients $\lambda_1$ and $\lambda_2$:
\begin{align}
\min _{\lambda_1, \lambda_2} 
& \sum_{\mathbf{x}_i \in \mathcal{D}_t} 
H\left(\mathcal{F}_{\Theta_\text{Cocktail}}\left(\mathbf{x}_i\right)\right), \\
& \text{where } \Theta_{\text{Cocktail}} = 
\Theta_{\text{pre}} + (\lambda_1\Delta\Theta_g) \oplus (\lambda_2\Delta\Theta_s).
\end{align}
For the LLM, the output of $\mathcal{F}_{\Theta_\text{Cocktail}}$ is a sentence. Since our instruction is to select a title from a given candidate set, the first few tokens output by the model are more important because after deciding on them, the subsequent tokens are more certain.
So in practice, we can only calculate the average entropy of the first few tokens in the sentence to represent $H\left(\mathcal{F}_{\Theta_\text{Cocktail}}\right)$. Good performance can be achieved with just 3 tokens.
Besides, we do not need all test data to be available. Even with only 50 unlabeled tests data, our method can have significant performance improvements.

\subsection{Discussion}
\noindent$\bullet$ \textbf{Key Advantages of \shortname.} 
\underline{1) Generalization.}
%\noindent$\bullet$\quad\textbf{Generalization.}
\shortname~is generalizable to various recommendation scenarios as it can adaptively determine how to fuse domain-general and domain-specific knowledge based on the distribution of the test data. Even in extreme cases where no training data is available for the new domain, \shortname~can still work using the generalizable base spirit.
Additionally, leveraging the in-context learning capabilities of LLMs, \shortname~naturally exhibits task generalization. E.g., it can generate explainable recommendation results.
\underline{2) Efficiency.}
%\noindent$\bullet$\quad\textbf{Efficiency.}
After adjusting the merging coefficients using a few unlabeled test data, we retains only a single LoRA module. As a result, there is no additional memory or computational overhead during inference.
\shortname~offers plug-and-play integration, where the domain-general module is trained once, and the domain-specific plugin is incorporated with minimal fine-tuning.

\noindent$\bullet$ \textbf{Comparison to Existing Methods.}
Traditional sequential recommendation models (e.g., GRU4Rec~\cite{DBLP:journals/corr/HidasiKBT15}, SASRec~\cite{kang2018self}) typically rely on explicit item IDs for modeling, limiting new domains or platforms generalization. \underline{Transferable methods} address this by unifying multi-domain data in the input space. E.g., VQ-Rec~\cite{10.1145/3543507.3583434} and UniSRec~\cite{hou2022towards} use text representations with contrastive pre-training for enhanced transferability.
With LLMs' emergence, \underline{breadth-oriented methods} leverage their generalization capabilities. P5~\cite{geng2022recommendation} unifies five recommendation tasks through a text-to-text paradigm. These methods enhance generalization by aggregating data from multiple domains/tasks. In contrast, \shortname~takes a parameter merging approach, integrating knowledge directly in the parameter space—offering greater flexibility and scalability.
\underline{Depth-oriented methods} focus on domain-specific alignment. TallRec~\cite{bao2023tallrec} uses LoRA for efficient adaptation, while LLaRA~\cite{10.1145/3626772.3657690}, iLoRA~\cite{kong2024customizing}, and AlphaRec~\cite{sheng2025language} align collaborative signals with LLMs. Although this significantly enhances performance in warm-start scenarios, it reduces generalization for new domains and cold-start scenarios.

\iffalse
\subsubsection{Model Generalization and Efficiency}
The proposed \shortname\newline paradigm is designed for ease of use, offering plug-and-play integration, where the domain-general module is trained once, and the domain-specific plugin is incorporated with minimal fine-tuning. \shortname~is highly efficient, as the mixture-of-LoRA method merges the weights, incurring no additional memory or computational cost during inference.
\noindent$\bullet$\quad\textbf{Generalization.}
\shortname~is generalizable to various recommendation scenarios.
For example, when facing a new recommendation domain, it only needs to train a new domain-specific LoRA module, enabling rapid generalization. Even in extreme cases where no training data is available for the new domain, \shortname~can still work using the generalizable base model. For the new user or new item recommendation scenario, \shortname~dynamically balances generalization and domain-specific specialization to deliver accurate recommendations.
Additionally, leveraging the in-context learning capabilities of LLMs, \shortname~naturally exhibits task generalization. For example, it can generate explainable recommendation results.

\noindent$\bullet$\quad\textbf{Efficiency.}
\shortname~is highly efficient, as the mixture-of-LoRA method merges the weights, incurring no additional memory or computational cost during inference.
\fi

%\subsubsection{Comparisons to LLM-Based Recommendation Models}

%% file: sections/experiments.tex
\begin{table*}[h]
\small
\centering
\caption{Performance Comparison in Warm Start I.I.D Scenario (Beauty, Toys, Sports, and MovieLens-1M).}
%\vspace{-2mm}
\begin{adjustbox}{max width=\textwidth}
\begin{tabular}{ll|cc|cc|cc||cc}
\toprule
\multirow{2}{*}{\textbf{Methods}} & \multirow{2}{*}{} & \multicolumn{2}{c|}{\textbf{Beauty}} & \multicolumn{2}{c|}{\textbf{Toys}} & \multicolumn{2}{c||}{\textbf{Sports}} & \multicolumn{2}{c}{\textbf{MovieLens-1M}} \\ 
 & & \textbf{NDCG@1} & \textbf{NDCG@3} & \textbf{NDCG@1} & \textbf{NDCG@3} & \textbf{NDCG@1} & \textbf{NDCG@3} & \textbf{NDCG@1} & \textbf{NDCG@3} \\ 
\midrule
\multirow{4}{*}{\textbf{Traditional}} 
& \textbf{BPR-MF} &  0.1630 & 0.2588 & 0.1276 & 0.2056 & 0.1496 & 0.2338 & 0.1724 & 0.4185 \\ 
& \textbf{GRU4Rec} & 0.1672 & 0.2752 & 0.1320 & 0.2243 & 0.1787 & 0.2829 & 0.1724 &  0.4423\\ 
& \textbf{SASRec} & 0.2410 & 0.3284 & 0.2223 & 0.3105 & 0.1957 & 0.2967 & 0.2257 & 0.4708 \\ 
& \textbf{FMLP-Rec}& 0.2988 & 0.4000 & 0.2994 & 0.3990 & 0.2645 & 0.3812 & 0.2410 & 0.5515 \\ 
\midrule
\multirow{2}{*}{\textbf{Transferable}} 
& \textbf{UniSRec} & 0.2654 & 0.4089 & 0.2612 & 0.3998 & 0.2341 & 0.3721 & 0.2615 & 0.5594 \\ 
& \textbf{VQ-Rec} & 0.2714 & 0.4157 & 0.2715 & 0.4119 & 0.2476 & \underline{0.3944} & 0.2805 & 0.5745 \\ 
\midrule
\multirow{5}{*}{\textbf{LLM-Based}} 
& \textbf{Qwen2-7B-zeroshot} & 0.0300 & 0.0394 & 0.0843 & 0.1062 & 0.0170 & 0.0242 & 0.0814 & 0.1057  \\ 
& \textbf{RecFormer} & 0.2858 & 0.3840 & 0.3001 & 0.3880 & 0.2667 & 0.3885 & 0.2743 & 0.5701 \\ 
& \textbf{P5} & 0.1775 & 0.2482 & 0.1171 & 0.1709 & 0.1860 & 0.2674 & 0.2046 & 0.2947\\ 
& \textbf{TALLRec} & 0.3347 & 0.3593 & 0.3746 & 0.3993 & 0.3585 & 0.3826 & 0.5392 & 0.5661\\ 
& \textbf{AlphaRec} & 0.2489 & 0.3456 & 0.2264 & 0.3193 & 0.2418 & 0.3538 & 0.2318 & 0.3661 \\ 
\midrule
\multirow{3}{*}{\textbf{Ours}} 
& \textbf{RecCocktail-WA} & 0.3722 & 0.3959 & 0.3722 & 0.3961 & 0.3410 & 0.3613 & \underline{0.5738} & \underline{0.5982} \\
& \textbf{RecCocktail-G} & 0.3081 & 0.3316 & 0.2957 & 0.3209 & 0.2750 & 0.2998 & 0.5680 & 0.5918 \\
& \textbf{RecCocktail-S} & \underline{0.4079} & \underline{0.4291} & \underline{0.4076} & \underline{0.4314} & \underline{0.3735} & 0.3925 & 0.5460 & 0.5703\\
& \cellcolor[gray]{0.9}\textbf{RecCocktail} 
& \cellcolor[gray]{0.9}\textbf{0.4132*} 
& \cellcolor[gray]{0.9}\textbf{0.4350*} 
& \cellcolor[gray]{0.9}\textbf{0.4097*} 
& \cellcolor[gray]{0.9}\textbf{0.4334*} 
& \cellcolor[gray]{0.9}\textbf{0.3754*} 
& \cellcolor[gray]{0.9}\textbf{0.3944*} 
& \cellcolor[gray]{0.9} \textbf{0.5783*} 
& \cellcolor[gray]{0.9} \textbf{0.6023*}\\ 
\bottomrule
\end{tabular}
\end{adjustbox}
\label{table:performance_comparison}
\end{table*}

\begin{table*}[h]
\small
\centering
\caption{Performance Comparison in Cold-Start Item O.O.D Scenario~(Beauty, Toys, Sports, and MovieLens-1M).}
%\vspace{-2mm}
\begin{adjustbox}{max width=\textwidth}
\begin{tabular}{ll|cc|cc|cc||cc}
\toprule
\multirow{2}{*}{\textbf{Methods}} & \multirow{2}{*}{} & \multicolumn{2}{c|}{\textbf{Beauty}} & \multicolumn{2}{c|}{\textbf{Toys}} & \multicolumn{2}{c||}{\textbf{Sports}} & \multicolumn{2}{c}{\textbf{MovieLens-1M}} \\ 
 & & \textbf{NDCG@1} & \textbf{NDCG@3} & \textbf{NDCG@1} & \textbf{NDCG@3} & \textbf{NDCG@1} & \textbf{NDCG@3} & \textbf{NDCG@1} & \textbf{NDCG@3} \\ 
\midrule
\multirow{4}{*}{\textbf{Traditional}} 
& \textbf{BPR-MF} & 0.0306 & 0.0688 & 0.0333 & 0.0765 & 0.0350 & 0.0739 & 0.0723 & 0.1421 \\ 
& \textbf{GRU4Rec} & 0.0562 & 0.1063 & 0.0447 & 0.0926 & 0.0640 & 0.0996 & 0.0798 & 0.1489 \\ 
& \textbf{SASRec} & 0.0656 & 0.1368 & 0.0670 & 0.1210 & 0.0547 & 0.1203 & 0.0912 & 0.1891 \\ 
& \textbf{FMLP-Rec} & 0.0587 & 0.1229 & 0.0537 & 0.1117 & 0.0545 & 0.1236 & 0.1145 & 0.1947 \\ 
\midrule
\multirow{2}{*}{\textbf{Transferable}} 
& \textbf{UniSRec} & 0.0957 & 0.1457 & 0.0814 & 0.1559 & 0.0832 & 0.1408 & 0.0985 & 0.1343 \\ 
& \textbf{VQ-Rec} & 0.1189 & 0.1589 & 0.0957 & 0.1603 & 0.0985 & 0.1463 & 0.1025 & 0.1412\\ 
\midrule
\multirow{5}{*}{\textbf{LLM-Based}} 
& \textbf{Qwen2-7B-zeroshot} & 0.0187 & 0.0260 & 0.0293 & 0.0356 & 0.0213 & 0.0273 & 0.0318 & 0.0407 \\ 
& \textbf{RecFormer} & 0.1051 & 0.1687 & 0.0913 & 0.1592 & 0.0922 & 0.1489 & 0.1108 & 0.1547\\ 
& \textbf{P5} & 0.0871 & 0.1466 & 0.0755 & 0.1358 & 0.0758 & 0.1355 & 0.0957 & 0.1319 \\ 
& \textbf{TALLRec} & 0.1480 & 0.1783 & 0.1624 & 0.1831 & 0.1251 & 0.1524 & 0.1458 & 0.1668 \\ 
& \textbf{AlphaRec} & 0.0588 & 0.1235 & 0.0553 & 0.1223 & 0.0565 & 0.1358 & 0.0923 & 0.1981\\ 
\midrule
\multirow{3}{*}{\textbf{Ours}} 
& \textbf{RecCocktail-WA} & \underline{0.1766} & \underline{0.2077} & \underline{0.1589} & \underline{0.1810} & \underline{0.1610} & \underline{0.1903} & \underline{0.1636} & \underline{0.2597} \\
& \textbf{RecCocktail-G} & 0.1746 & 0.2072 & 0.1474 & 0.1710 & 0.1581 & 0.1868 & 0.1455& 0.1890\\
& \textbf{RecCocktail-S} & 0.1603 & 0.1863 & 0.1504 & 0.1764 & 0.1564 & 0.1867 & \underline{0.1636} & \underline{0.2597}\\
& \cellcolor[gray]{0.9}\textbf{RecCocktail}
& \cellcolor[gray]{0.9}\textbf{0.1825*} 
& \cellcolor[gray]{0.9}\textbf{0.2145*} 
& \cellcolor[gray]{0.9}\textbf{0.1614*} 
& \cellcolor[gray]{0.9}\textbf{0.1821*} 
& \cellcolor[gray]{0.9}\textbf{0.1646*} 
& \cellcolor[gray]{0.9}\textbf{0.1921*} 
& \cellcolor[gray]{0.9}\textbf{0.1818*} 
& \cellcolor[gray]{0.9}\textbf{0.2755*} \\ 
\bottomrule
\end{tabular}
\end{adjustbox}
\label{table:performance_comparison_cold_start}
\end{table*}

\section{Experiments}
\iffalse
In this section, we conduct experiments on three real-world datasets to answer the following research questions: 
\begin{itemize}[leftmargin=*]
\item \textbf{RQ1:} How does our proposed \shortname~perform compared to traditional, transferable, and LLM-based recommenders? How does \shortname~ perform in terms of generalization in scenarios such as the cold-start setting?
\item \textbf{RQ2:} How does each component of \shortname~ affect its effectiveness? Can the mixture-of-LoRA method integrate domain-specific and domain-general knowledge effectively?
\item \textbf{RQ3:} Is MoLoRec robust to hyperparameter selection and instruction input?
\end{itemize}
\fi
\subsection{Experimental Settings}
\iffalse
\begin{table}[ht]
\small
\centering
\caption{Statistics of the datasets for stage 1.}
\begin{tabular}{lcccc}
\toprule[1pt]  % Make the top border thicker
\textbf{Datasets} & \textbf{\# Users} & \textbf{\# Items} & \textbf{\# Interactions} & \textbf{Density(\%)} \\
\midrule[0.8pt]     % Optional: Adjust midrule thickness if needed
Clothing & 39,387 & 23,033 & 278,677 & 0.0307 \\
Cell & 27,879 & 10,429 & 194,439 & 0.0669 \\
Grocery & 14,681 & 8,713 & 151,254 & 0.1182 \\
Health & 38,609 & 18,534 & 346,355 & 0.0484 \\
Home & 66,519 & 28,237 & 551,682 & 0.0294 \\
Pet & 19,856 & 8,510 & 157,836 & 0.0934 \\
Tools & 16,638 & 10,217 & 134,476 & 0.0791 \\
Videos & 24,303 & 10,672 & 231,780 & 0.0894 \\
\midrule
Total & 247,872 & 118,354 & 2,046,499 & - \\
\bottomrule[1pt]  % Make the bottom border thicker
\end{tabular}
\label{tab:stage1_data}
\end{table}
\fi

\iffalse
\begin{table}[ht]
\small
\centering
\caption{Statistics of the four datasets.}
\begin{tabular}{lcccc}
\toprule[1pt]  % Make the top border thicker
\textbf{Specific Datasets} & \textbf{\# Users} & \textbf{\# Items} & \textbf{\# Interactions} & \textbf{Density(\%)} \\
\midrule[0.8pt]     % Optional: Adjust midrule thickness if needed
Beauty & 22,363 & 12,101 & 198,502 & 0.0734 \\
Toys & 19,412 & 11,924 & 167,597 & 0.0724 \\
Sports & 35,598 & 18,357 & 296,337 & 0.0453 \\
Movielens-1M & 6,040 & 6,883 & 1,000,209 & 0.2410 \\
\bottomrule[1pt]  % Make the bottom border thicker
\end{tabular}
\label{tab:data}
\end{table}
\fi
\noindent$\bullet$ \textbf{Datasets.}
%\subsubsection{\textbf{Datasets.}}
We conduct experiments on e-commerce and movie recommendation scenarios. 
For the e-commerce recommendation scenario, the domain-general instruction tuning dataset is conducted using seven e-commerce domains in Amazon\footnote{\url{https://jmcauley.ucsd.edu/data/amazon/.}} and validated on three domain-specific datasets in Amazon~(Beauty, Toys, Sports). 
For the movie recommendation scenario, the domain-general dataset is built using MovieLens-10M\footnote{\url{https://grouplens.org/datasets/movielens/}} and validated on the domain-specific dataset MovieLens-1M.

For all datasets, items are represented using their textual ``title" information. To prevent data leakage, we carefully removed the overlapping portions between the domain-general dataset and the domain-specific datasets. We consider two recommendation settings:
1) \textbf{Warm-Start Setting} keeps the five-core dataset and filters users and items with fewer than five interactions for all datasets. Following \cite{geng2022recommendation,lin2024bridging}, we adopt the leave-one-out strategy to split the filtered dataset. More concretely, we split the last interaction of each user into the test set, the second-to-last one into the validation set, and the rest into the training data. 
2) \textbf{New-Item Setting} uses the same training and validation sets as the warm-start setting, but replaces the items in the test set with those that never appear in the training or validation sets as ground-truth.

\noindent$\bullet$ \textbf{Baselines.}
%\subsubsection{\textbf{Baselines.}} 
Baselines include traditional recommendation methods (BPR-MF~\cite{10.5555/1795114.1795167}, GRU4Rec~\cite{DBLP:journals/corr/HidasiKBT15}, SASRec~\cite{kang2018self}, and FMLP-Rec)~\cite{10.1145/3485447.3512111}, transferable sequential recommenders (UniSRec~\cite{hou2022towards}, VQ-Rec~\cite{10.1145/3543507.3583434}), LLM-based recommenders (Qwen2-7B-zeroshot~\cite{qwen2}, RecFormer~\cite{10.1145/3580305.3599519}, P5~\cite{geng2022recommendation}, TALLRec~\cite{bao2023tallrec}, AlphaRec~\cite{sheng2025language}) and our ablation counterparts (\shortname-WA, \shortname-G, \shortname-S).

\noindent$\bullet$ \textbf{Evaluation Setting}
\label{sec:evluation_setting}
Following LLM-based recommendation works~\cite{10.1145/3708882,kim2024large}, we add 29 randomly selected non-interacted items to the candidate set, so that for each user it contains 1 positive item and 29 negative items. For quantitative comparison, we employ 
widely used ranking-based metrics, NDCG@1 and NDCG@3 for all experiments. All metrics are ``the higher, the better". For all tables in the following, \textbf{bold*} numbers refer to the best performance, while \underline{underlined} numbers indicate the second-best performance.

\noindent\textbf{(See Appendix \ref{sec:appendix_dataset} for dataset statistics, Appendix \ref{sec:appendix_baselines} for more baseline details, and Appendix \ref{sec:appendix_implementation} for implementation details.)}

\begin{figure*}[ht]
    \centering
    \begin{subfigure}[t]{0.245\textwidth}
        \centering
        \includegraphics[width=1\textwidth,height=0.15\textheight]{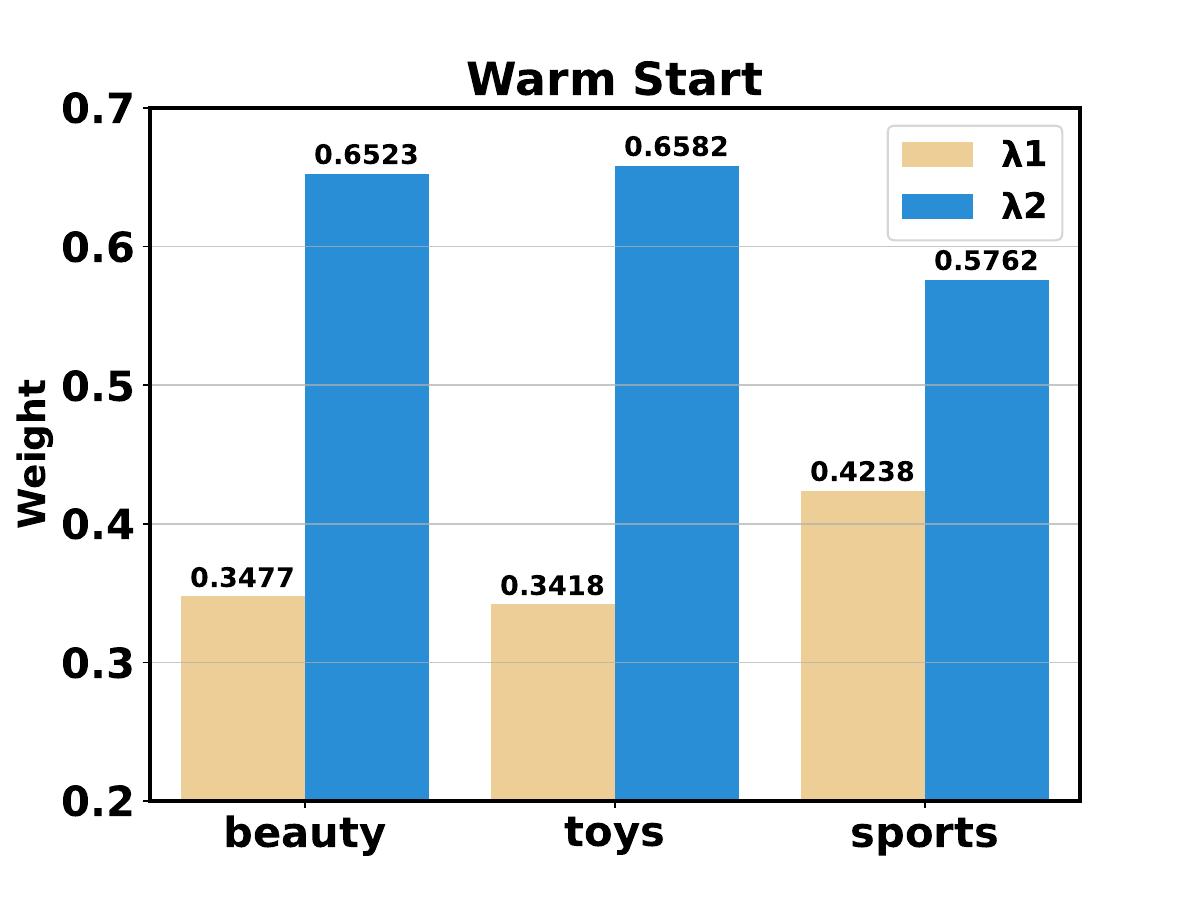}
        \caption{Warm-Start Scenario}
        \label{fig:coeff_1}
    \end{subfigure}
    \hfill
    \begin{subfigure}[t]{0.245\textwidth}
        \centering
        \includegraphics[width=1\textwidth,height=0.15\textheight]{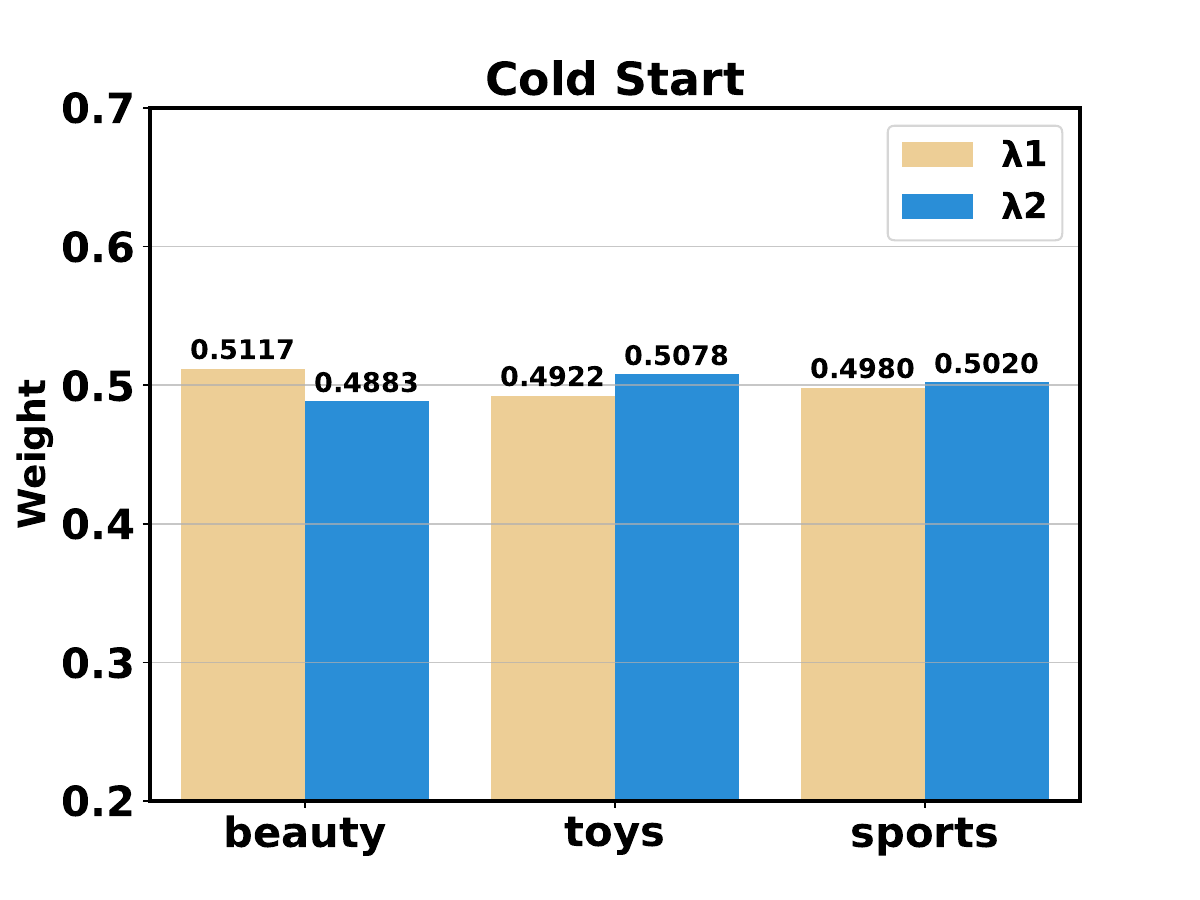}
        \caption{Cold-Start Scenario}
        \label{fig:coeff_2}
    \end{subfigure}
    \hfill
    \begin{subfigure}[t]{0.245\textwidth}
        \centering
        \includegraphics[width=1\textwidth,height=0.142\textheight]{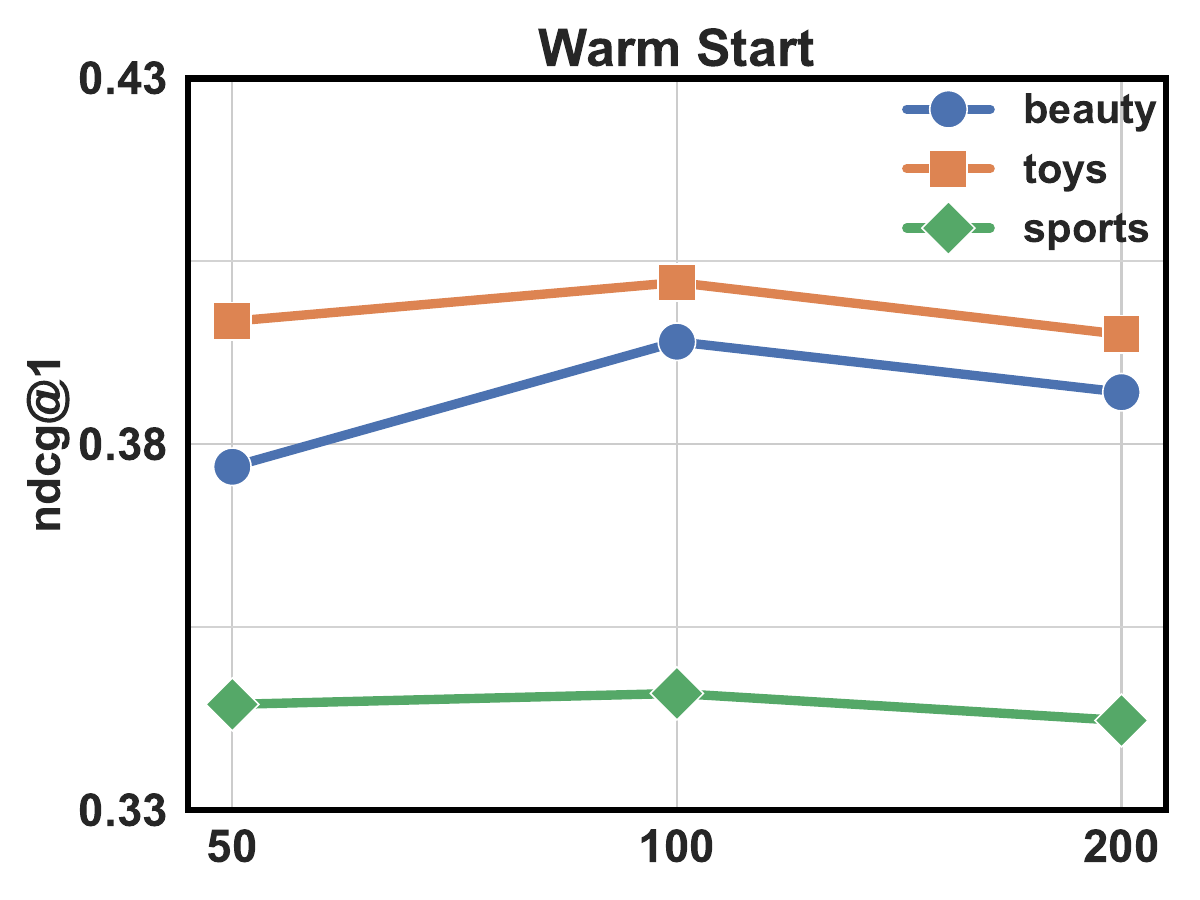}
        \caption{Warm-Start Scenario}
        \label{fig:111}
    \end{subfigure}
    \hfill
    \begin{subfigure}[t]{0.245\textwidth}
        \centering
        \includegraphics[width=1\textwidth,height=0.142\textheight]{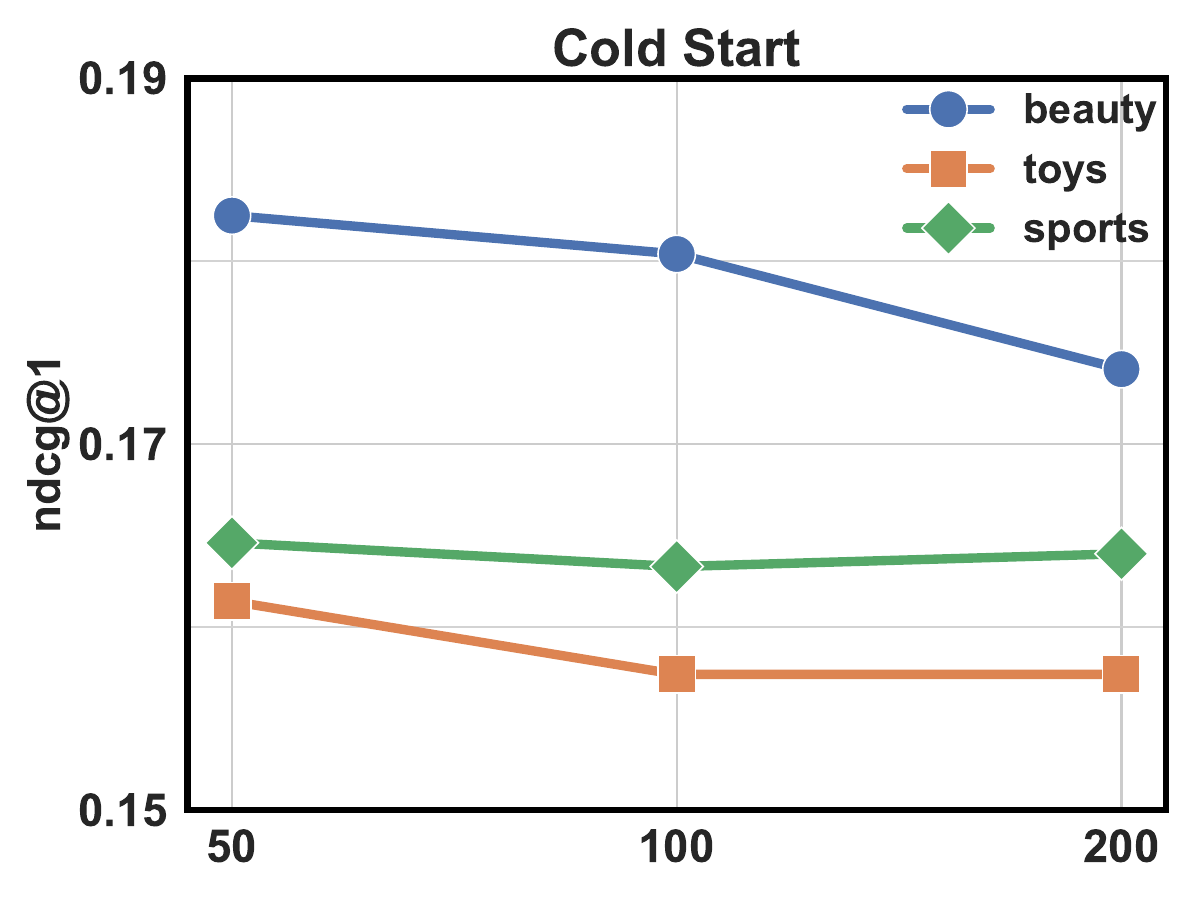}
        \caption{Cold-Start Scenario}
        \label{fig:222}
    \end{subfigure}
    \caption{(a) and (b) are the coefficients $\lambda_1$ and $\lambda_2$ calculated by entropy-guided adaptive merging. (c) and (d) are the impact of the number of unlabeled test data.}
    \label{fig:ts}
\end{figure*}

\subsection{Overall Performance}
\label{sec:overall_performance}
We comprehensively compare \shortname~against traditional, transferable, and LLM-based recommenders. 
Tables \ref{table:performance_comparison} and \ref{table:performance_comparison_cold_start} present results for warm-start I.I.D and new-item O.O.D settings, respectively. Results indicate that:
\begin{itemize}[leftmargin=*]
    \item The proposed \shortname~consistently achieves the best performance across all I.I.D and O.O.D scenarios on the four datasets, with a t-test at p$<$0.05 level. The strong performance of \shortname~demonstrates its ability to efficiently capture domain-specific knowledge while maintaining excellent generalization capabilities.
    %Specifically, in the warm-start scenario, \shortname~ achieves notable improvements in NDCG@1 over the best baseline methods (excluding our ablation counterparts), with performance gains of 28.8\%, 23.85\%, 25.05\%, and 21.52\% across the Beauty, Toys, Sports, and Movielens-1M, respectively. In the cold-start O.O.D scenario, the improvements are 28.98\%, 29.02\%, 34.26\%, and 24.69\%. The outstanding performance of \shortname~in both I.I.D and O.O.D scenarios demonstrates its ability to efficiently capture domain-specific knowledge while exhibiting exceptional generalization capabilities.
    %\item Qwen2-7B demonstrates limited performance across all scenarios. However, the LLMs trained via TALLRec achieved significant improvements. This is because there is a gap between pre-training general text corpus of LLMs and the recommendation task, showing the importance of using recommendation knowledge for instruction fine-tuning on pre-trained LLMs.
    \item Traditional recommendation methods and the ID-based LLM recommendation method P5, AlphaRec perform poorly in cold-start scenarios. Relying heavily on collaborative filtering information reduces the model's generalization capability.
    %\item Among traditional recommenders, sequential methods (GRU4Rec, SASRec, FMLP-Rec) surpass non-sequential methods (BPR-MF) on 4 datasets in the warm-start scenario. The better performance stems from the sequential modeling of the user’s interaction sequence, which captures dynamic shifts in user interests and intricate item dependencies. We find that TALLRec, when using only textual information, outperforms traditional recommendation methods, highlighting the potential of large language model-based recommendation approaches.
    \item We observe that \shortname~consistently outperforms its ablation counterparts across all scenarios. This highlights the importance of integrating both general recommendation knowledge and domain-specific insights, which are complementary. The results also confirm the effectiveness of our LoRA cocktail method for knowledge fusion. Note that the \shortname-G has not been exposed to training data from the Beauty, Toys, Sports, or Movielens-1M domains. It still achieves commendable performance in such zero-shot settings, demonstrating that it has effectively learned generalizable recommendation knowledge. We also find that simple average merging (\shortname-WA) sometimes trigger negative transfer, demonstrating the effectiveness and importance of our proposed Entropy-Guided Adaptive Merging strategy for selecting appropriate merging coefficients.
\end{itemize}

\subsection{In-Depth Analysis}
\iffalse
\subsubsection{\textbf{Ablation Study}}
From the performance of \shortname, \shortname-G, \shortname-S in the Table \ref{table:performance_comparison} and \ref{table:performance_comparison_cold_start}, we can find that:
1) \shortname~ consistently surpasses its ablation counterparts in all scenarios, which indicates that it is crucial for LLM to comprehend both general recommendation world knowledge and domain-specific knowledge. These two types of knowledge can complement each other. It also validates that the LoRA cocktail method can effectively integrate both types of knowledge. 2) Note that the \shortname-G has not been exposed to training data from the Beauty, Toys, Sports, or Movielens-1M domains in any experimental setting. However, it still achieves commendable performance in such zero-shot settings, demonstrating that it has effectively learned generalizable knowledge in the recommendation domain. This strong performance underscores its robust generalization ability.
\fi

\noindent$\bullet$ \textbf{Analysis of the Coefficients $\lambda_1$ and $\lambda_2$.}
In Figure \ref{fig:coeff_1} and \ref{fig:coeff_2}, we investigate the coefficients $\lambda_1$ and $\lambda_2$ calculated by entropy-guided adaptive LoRA cocktail in the warm-start scenario and the cold-start scenario, respectively. $\lambda_1$ and $\lambda_2$ represent the respective weights assigned to the base spirit and the ingredient LoRA module during their merging. We observe that in the warm-start scenario, $\lambda_2$ is relatively large, reflecting a greater reliance on domain-specific LoRA. Conversely, in the cold-start scenario, the weight of $\lambda_1$ increases significantly, emphasizing the importance of domain-general knowledge. This result is reasonable and aligns with the differing requirements of these two scenarios. 
This result also demonstrates the effectiveness of our entropy-guided adaptive LoRA cocktail method.

\noindent$\bullet$ \textbf{Analysis of the Number of Unlabeled Test Data.}
The number of unlabeled test data is one of the hyperparameters. Figure \ref{fig:111} and \ref{fig:222} illustrates the impact of different numbers on the NDCG@1 performance. The entropy-guided learning method converges rapidly, experimental results indicate that setting the number to 50 or 100 achieves a model fusion weight with optimal performance, this configuration proves effective across the majority of experiments conducted on the Beauty, Toys, Sports and MovieLens-1M datasets.
Therefore, a limited amount of unlabeled test data is sufficient to learn suitable model fusion weights, thereby significantly reducing the data requirements and computation resource costs.

\noindent$\bullet$ \textbf{Performance in Few-Shot Training Setting.}
% \begin{wraptable}{r}{0.65\textwidth}
% \centering
% % \vspace{-1.1em}
% \small
% \caption{NDCG@1 performance in the few-shot training setting on Movielens-1M Dataset.}
% \label{tab:few-shot}
% %\vspace{-1em}
% \begin{tabular}{ccccc}
% \toprule
% \textbf{Scenario} & \textbf{Sample} & \textbf{TallRec} & \textbf{2nd Finetune} & \textbf{RecCocktail} \\
% \midrule
% \multirow{3}{*}{\textbf{Warm}} 
% & \textbf{10\%}  & 0.3957  & 0.0704 & 0.5353 \\
% & \textbf{20\%}  & 0.4298  & 0.0790 & 0.5454 \\
% & \textbf{30\%}  & 0.4563  & 0.0540 & 0.5498 \\
% \midrule
% \multirow{3}{*}{\textbf{Cold}} 
% & \textbf{10\%}  & 0.1091  & 0.0000 & 0.1455 \\
% & \textbf{20\%}  & 0.1091  & 0.0182 & 0.1636 \\
% & \textbf{30\%}  & 0.1273  & 0.0000 & 0.1636 \\
% \bottomrule
% \end{tabular}
% %\vspace{-1em}
% \end{wraptable}
\begin{table}[!t]
    \centering
    \small
    \caption{NDCG@1 performance in the few-shot training setting on Movielens-1M Dataset.}
    \label{tab:few-shot}
    \begin{tabular}{ccccc}
    \toprule
    \textbf{Scenario} & \textbf{Sample} & \textbf{TallRec} & \textbf{2nd Finetune} & \textbf{RecCocktail} \\
    \midrule
    \multirow{3}{*}{\textbf{Warm}} 
    & \textbf{10\%}  & 0.3957  & 0.0704 & 0.5353 \\
    & \textbf{20\%}  & 0.4298  & 0.0790 & 0.5454 \\
    & \textbf{30\%}  & 0.4563  & 0.0540 & 0.5498 \\
    \midrule
    \multirow{3}{*}{\textbf{Cold}} 
    & \textbf{10\%}  & 0.1091  & 0.0000 & 0.1455 \\
    & \textbf{20\%}  & 0.1091  & 0.0182 & 0.1636 \\
    & \textbf{30\%}  & 0.1273  & 0.0000 & 0.1636 \\
    \bottomrule
    \end{tabular}
\end{table}
We further conduct experiments in scenarios with limited domain-specific training data. Specifically, we adopt a few-shot training setup on MovieLens-1M, where only a small percentage of samples are randomly selected from the training set for model training.
We compare \shortname~with TallRec and the results of second-round fine-tuning on the generalizable base model. 
The experimental results are shown in Table \ref{tab:few-shot}.
We find that the optimization approach of second-round fine-tuning led to catastrophic forgetting. It fails to generate output in the specified instruction format. The experimental results demonstrate that \shortname~maintains strong performance even in few-shot scenarios. 

\textbf{(See Appendix \ref{sec:appendix_experiments} for more experimental results.)}

%% file: sections/conclusion.tex
\vspace{-0.1cm}
\section{Conclusion}
In this paper, we proposed a generalizable and efficient LLM-based recommendation framework \shortname. As the recommendation data for a specific domain is limited, \shortname~ is designed to combine domain-general and domain-specific recommendation knowledge. Specifically, \shortname~first construct domain-general LoRA as base spirit and domain-specific LoRA as ingredient. Then the combination is achieved by merging two fine-tuning model parameters. \shortname~ is generalizable as it injects recommendation general knowledge to any domain-specific recommendation tasks. Besides, \shortname~ is efficient not only because it chooses parameter-efficient fine-tuning, but also the plug-and-play nature of any domain-specific recommendation task and domain-general task. Extensive experimental results on four recommendation datasets under both the warm scenario and cold-start scenario show the effectiveness of our proposed framework.
%In the future, we intend to adapt MoLoRec framework to the scenarios of multi-task and multi-domain recommendation. 

%the following aspects. On one hand, we would like to diversify instruction data with multiple related tasks for recommendation, such as  user profiling and review summarization. 

%To that end, in this paper, we propose a generalizable and efficient LLM-based recommendation framework. Our approach starts by parameter efficient tuning a domain-general module with general recommendation instruction data, such to align LLM with recommendation task. Then, given users' behavior of a specific domain, we construct a domain-specific instruction dataset and apply efficient fine-tuning to the pre-trained LLM. After that, we provide approaches to integrate the above domain-general parts and domain-specific parts with mixture of parameters from two parts. Please note that, our proposed framework is efficient with plug and play, as domain-general module is trained only once, and any domain-specific plug-in can be efficiently merged with only domain-specific fine-tuning. Extensive experiments on multiple datasets validate the effectiveness and generality of our proposed framework. Our codes are available at \url{https://anonymous.4open.science/r/MoLoRec}.

%% file: sections/appendix.tex
\newpage
\section{Experimental Details}
\label{sec:appendix_experiments}
\begin{table}[!t]
\centering
\caption{Statistics of the datasets for stage 1 and specific domains.}
\scalebox{0.8}{ 
\begin{tabular}{lcccc}
\toprule[1.6pt]  % Make the top border thicker
%\midrule[0.8pt]
\multicolumn{5}{c}{\textbf{Domain-General Dataset-1}} \\
\midrule[0.8pt]
\textbf{Datasets} & \textbf{\# Users} & \textbf{\# Items} & \textbf{\# Interactions} & \textbf{Density(\%)} \\
\midrule[0.8pt]
Clothing & 39,387 & 23,033 & 278,677 & 0.0307 \\
Cell & 27,879 & 10,429 & 194,439 & 0.0669 \\
Grocery & 14,681 & 8,713 & 151,254 & 0.1182 \\
Health & 38,609 & 18,534 & 346,355 & 0.0484 \\
Home & 66,519 & 28,237 & 551,682 & 0.0294 \\
Pet &

19,856 & 8,510 & 157,836 & 0.0934 \\
Tools & 16,638 & 10,217 & 134,476 & 0.0791 \\
Videos & 24,303 & 10,672 & 231,780 & 0.0894 \\
%\midrule
%Total & 247,872 & 118,354 & 2,046,499 & - \\
\midrule[0.8pt]
\multicolumn{5}{c}{\textbf{Domain-Specific Dataset-1}} \\
\midrule[0.8pt]
Beauty & 22,363 & 12,101 & 198,502 & 0.0734 \\
Toys & 19,412 & 11,924 & 167,597 & 0.0724 \\
Sports & 35,598 & 18,357 & 296,337 & 0.0453 \\
\midrule[1.6pt]
%\midrule[0.8pt]
\multicolumn{5}{c}{\textbf{Domain-General Dataset-2}} \\
\midrule[0.8pt]
Movielens-10M & 71,567 & 10,681 & 10,000,054 & 1.3082 \\
\midrule[0.8pt]
\multicolumn{5}{c}{\textbf{Domain-Specific Dataset-2}} \\
\midrule[0.8pt]
Movielens-1M & 6,040 & 6,883 & 1,000,209 & 2.4059 \\
%\midrule[0.8pt]
\bottomrule[1.6pt]  % Make the bottom border thicker
\end{tabular}
}
\label{tab:data}
\end{table}

\subsection{Dataset Statistics}
\label{sec:appendix_dataset}
The statistics of the domain-general and the domain-specific datasets are shown in Table \ref{tab:data}.
\subsection{Baselines}
\label{sec:appendix_baselines} 
We compare \shortname~ with traditional recommendation methods (BPR-MF, GRU4Rec, SASRec, and FMLP-Rec), transferable sequential recommenders (UniSRec, VQ-Rec), LLM-based recommenders (Qwen2-7B-zeroshot, RecFormer, P5, TALLRec, AlphaRec) and our ablation counterparts (RecCocktail-G, RecCocktail-S). 
\begin{itemize}[leftmargin=*]
    \item \textbf{BPR-MF}~\cite{10.5555/1795114.1795167} is one of the most representative collaborative filtering models.
    \item \textbf{GRU4Rec}~\cite{DBLP:journals/corr/HidasiKBT15} is a seminal method that uses RNNs to model user action sequences for session-based recommendation. 
    \item \textbf{SASRec}~\cite{kang2018self} is a representative sequential recommender model that adopts a self-attention mechanism to learn the item dependency from user interactions.
    \item \textbf{FMLP-Rec}~\cite{10.1145/3485447.3512111} is an all-MLP model with learnable filters for sequential recommendation tasks.
    \item \textbf{UniSRec}~\cite{hou2022towards} equips textual item representations with an MoE-enhanced adaptor for domain fusion and adaptation. Both item-sequence and sequence-sequence contrastive learning tasks are designed for pre-training transferable sequence representations.
    \item \textbf{VQ-Rec}~\cite{10.1145/3543507.3583434} learns vector-quantized item representations for transferable sequential Recommenders.
    \item \textbf{Qwen2-7B-zeroshot}\footnote{\url{https://huggingface.co/Qwen/Qwen2-7B-Instruct}} is a well-known open-source LLM. In our experiments, we choose it as \shortname~'s LLM backbone.
    \item \textbf{RecFormer}~\cite{10.1145/3580305.3599519} models user preferences and item features using the LongFormer~\cite{beltagy2020longformer} backbone, transforming sequential recommendation into a task of predicting the next item as if predicting the next sentence, by converting item attributes into a sentence format.
    \item \textbf{P5}~\cite{geng2022recommendation} is a unified LLM-based recommendation framework. It is built on T5 by fine-tuning with multiple recommendation tasks.
    \item \textbf{TALLRec}~\cite{bao2023tallrec} learns the recommendation task based on prompts consisting solely of text and fine-tunes the LLMs using the LoRA. We use LLaMA-7B as the LLM backbone, following the original paper. For constructing instruction data, we adopt the same prompt format as us.
    \item \textbf{AlphaRec}~\cite{sheng2025language} integrates language representations for recommendation. It can adjust recommendations according to text-based user intentions, enabling recommenders to evolve into intention-aware systems through a straightforward paradigm shift.
    \item \textbf{RecCocktail-WA} is an ablation counterpart of our proposed framework. It uses Weight Averaging (WA) to merge domain-general and domain-specific LoRAs instead of the Entropy-Guided Adaptive Merging strategy.
    \item \textbf{RecCocktail-G} is an ablation counterpart of our proposed framework. It only underwent stage 1, utilizing only the domain-general LoRA module.
    \item \textbf{RecCocktail-S} is an ablation counterpart of our proposed framework. It only underwent stage 2, utilizing only the domain-specific LoRA module.
\end{itemize}

\subsection{Implementation Details}
\label{sec:appendix_implementation}
To ensure a fair comparison, the experimental settings are standardized as follows:
For traditional recommendation methods (BPR-MF, GRU4Rec, SASRec, and FMLP-Rec), the learning rate is set to 0.001, and the Adam optimizer is employed. The batch size is set to 256, and the embedding dimension is set to 64.
Regarding transferable sequential recommenders (UniSRec, VQ-Rec), these models utilize a BERT for text processing. Specifically, the pre-trained models provided by the original authors are fine-tuned on our dataset. For RecFormer, the pre-trained model provided by the original work is also fine-tuned on the downstream tasks.
In the case of P5 and TALLRec, an identical instruction fine-tuning template is used to align the original models with the recommendation tasks. Here, P5 undergoes full fine-tuning, while TALLRec is fine-tuned using a LoRA approach with a rank of 16.
This setup standardizes the evaluation framework across different recommendation methodologies, ensuring comparability and fairness in assessing their performance.
We use Qwen2-7B as the LLM backbone for \shortname. For PEFT conducted on  NVIDIA RTX 4090(24G) GPUs, we adopt LoRA with rank as 16, alpha as 32, and LoRA dropout as 0.05 to get general LoRA and specific LoRA. The learning rate is selected from \{1e-4, 2e-4\} and the batch size is set to 128. For LoRA adapters fusion weights learning conducted on two NVIDIA RTX 4090(24G) GPUs, the batch size is set to 60, the number of test samples for learning $\lambda_1$ and $\lambda_2$ is grid searched from \{50, 100\}. The number of tokens at the beginning of each title involved in training is set to 3 for Toys and Sports and 5 for Beauty because titles in Beauty are longer than Toys and Sports. In order to reduce GPU memory usage, we employed gradient checkpointing techniques. Since the model structure has not changed, we use the VLLM inference acceleration framework to perform inference and then evaluate the results. Other implementation details are available in our source code.

\iffalse
\subsection{\textbf{Evaluation Setting}} 
\label{sec:evluation_setting}
Following some previous LLM-based recommendation works~\cite{10.1145/3708882,kim2024large}, to evaluate the performance of the sequential recommendation models, we add 29 randomly selected non-interacted items to the test set, so that the test set of each user contains 1 positive item and 29 negative items. For quantitative comparison, we employ 
widely used ranking-based metrics, NDCG@1 and NDCG@3 for all experiments. All metrics are ``the higher, the better". For all tables in the following, \textbf{bold*} numbers refer to the best performance, while \underline{underlined} numbers indicate the second-best performance.
\fi
\subsection{\textbf{Experiments on More LLM Backbones}} 
To evaluate the robustness of the RecCocktail framework to different LLM backbones, we also utilize Llama-3.1-8B\footnote{\url{https://huggingface.co/meta-llama/Llama-3.1-8B}} as LLM backbone and observe the experimental results.
As shown in Table \ref{table:llama3}, when switching the LLM backbone to LLaMA-3.1-8B, \shortname~improves over both \shortname-G and \shortname-S on all datasets, indicating the effectiveness of adaptive merging.

\begin{table*}[h]
\small
\centering
\caption{Ablation Study on Llama-3.1-8b Across Datasets in Cold-Start Item O.O.D Scenario (Beauty, Toys, and Sports).}
\begin{adjustbox}{max width=\textwidth}
\begin{tabular}{l|cc|cc|cc}
\toprule
\textbf{Methods} & \multicolumn{2}{c|}{\textbf{Beauty}} & \multicolumn{2}{c|}{\textbf{Toys}} & \multicolumn{2}{c}{\textbf{Sports}} \\
 & \textbf{NDCG@1} & \textbf{NDCG@3} & \textbf{NDCG@1} & \textbf{NDCG@3} & \textbf{NDCG@1} & \textbf{NDCG@3} \\
\midrule
\textbf{RecCocktail-G} & \underline{0.1786} & \underline{0.2083} & \underline{0.1554} & 0.1721 & \underline{0.1538} & \underline{0.1814} \\
\textbf{RecCocktail-S} & 0.1663 & 0.1911 & 0.1534 & \underline{0.1723} & 0.1509 & 0.1809 
\\
\cellcolor[gray]{0.9}\textbf{RecCocktail} 
& \cellcolor[gray]{0.9}\textbf{0.1801*} 
& \cellcolor[gray]{0.9}\textbf{0.2086*} 
& \cellcolor[gray]{0.9}\textbf{0.1683*} 
& \cellcolor[gray]{0.9}\textbf{0.1882*} 
& \cellcolor[gray]{0.9}\textbf{0.1587*} 
& \cellcolor[gray]{0.9}\textbf{0.1864*} \\
\bottomrule
\end{tabular}
\end{adjustbox}
\label{table:llama3}
\end{table*}

\subsection{\textbf{Case Study}}
\begin{figure}[ht]
    \centering
    \includegraphics[width=0.95\columnwidth]{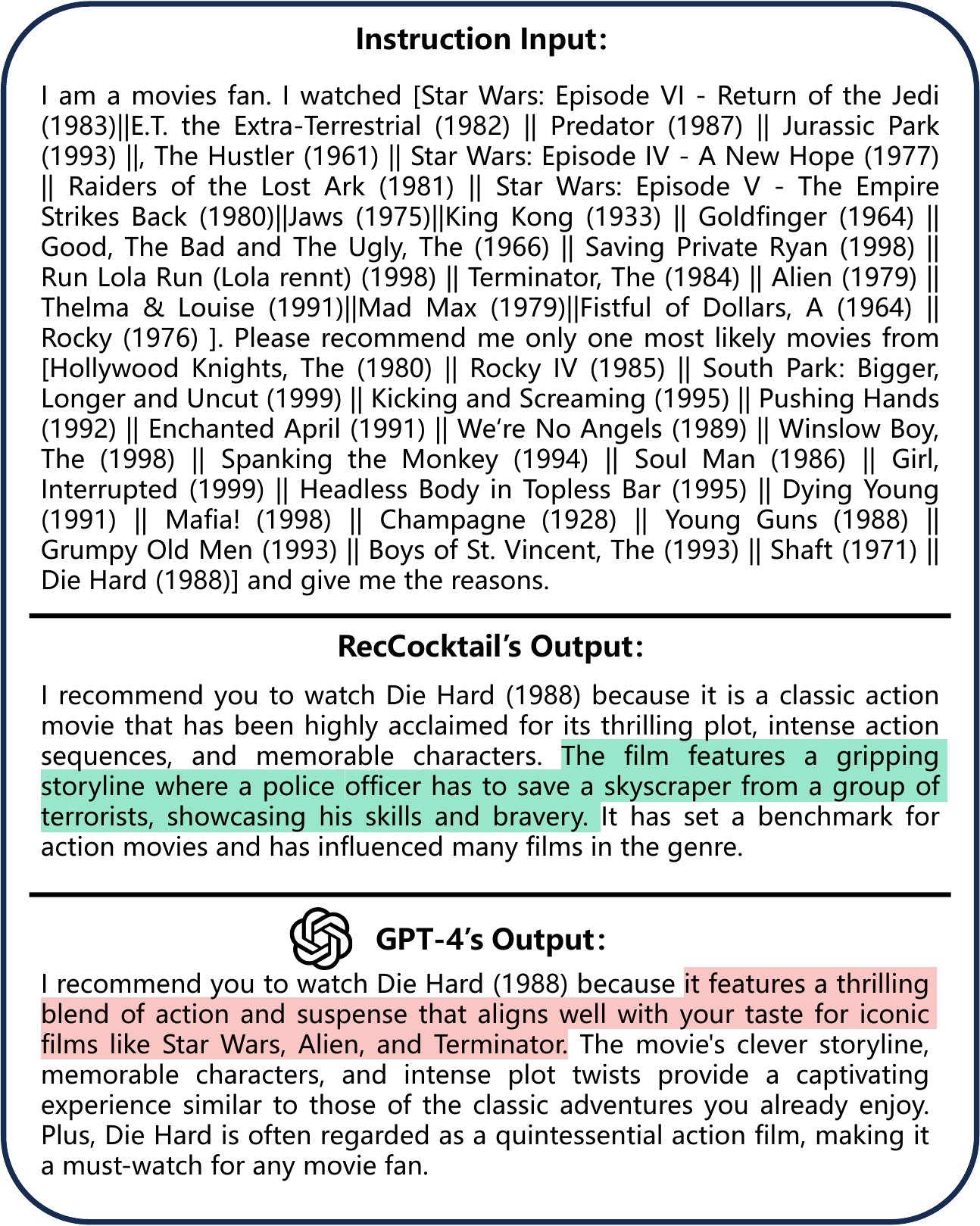}
    \caption{Case study of \shortname~ and GPT-4 explainable recommendation results.}
    \label{fig:case}
\end{figure}

We further conduct a case study to delve deeper into the recommendation results of \shortname. 
We randomly selected a user from the Movielens-1M test set, provided their historical viewing records and a candidate set, and asked both our model and ChatGPT to make movie recommendations along with explanations for their choices.
The outputs are shown in Figure \ref{fig:case}.
We find that \shortname~ successfully generalizes to the explainable recommendation task.
It accurately captured the user's preference for action movies from their historical viewing records and leveraged world knowledge to provide an accurate interpretation of the plot of Die Hard. 
In contrast, GPT-4 lacks domain-specific knowledge in the recommendation, incorrectly associating the action movie Die Hard with science fiction films like Star Wars, Alien, and Terminator, resulting in unreasonable explanation outcomes.
This case study further highlights \shortname's task generalization capability and its deep understanding of recommendation knowledge.

% \begin{figure}[ht]
%     \centering
%     \includegraphics[width=0.7\textwidth]{figures/case_4.pdf}
%     %\vspace{-0.4cm}
%     \caption{Case study of \shortname~ and GPT-4 explainable recommendation results.}
%     %\vspace{-0.4cm}
%     \label{fig:case}
% \end{figure}
% \input{sections/related}

\iffalse
\section{Limitation}
\label{sec:limitation}
While \shortname~demonstrates promising results, several limitations remain. First, due to computational constraints, we have not explored the use of larger LLM backbones (e.g., models with 30B parameters), which may further enhance recommendation quality. Second, our study does not thoroughly examine the impact of merging multiple domain-specific "ingredients" into the "base spirit" simultaneously. Finally, our framework is built upon LoRA, the most widely used parameter-efficient fine-tuning method, without evaluating alternative techniques such as prefix-tuning or adapters. These limitations point to valuable directions for future work.

\section{Broader Impact}
\label{sec:broader_impact}
Our work advances the field of LLM-based recommendation by demonstrating how large language models can be both generalized and specialized within a unified and efficient framework. By bridging the gap between domain-general and domain-specific recommendation paradigms, the proposed framework, \shortname, effectively supports diverse recommendation scenarios without incurring additional memory or latency overhead during inference. Beyond e-commerce applications, the core principles of \shortname~are broadly applicable to domains such as education, healthcare, and personalized content delivery. Overall, \shortname~ represents a meaningful step toward enhancing user experience and enabling more personalized and scalable recommendation solutions across a wide range of digital applications.
\fi

\input{sections/related}

%% file: sections/related.tex
\section{Related Work}
\noindent$\bullet$\quad\textbf{Sequential Recommendation.}
Sequential recommendation utilizes users' interaction histories to predict the next relevant item. Deep learning-based methods, such as RNNs~\cite{DBLP:journals/corr/HidasiKBT15,RNN2}, GNNs~\cite{graph_1,graph_2}, and attention mechanisms~\cite{kang2018self,S3-Rec,CL4SRec}, have become mainstream but rely solely on item IDs, limiting their adaptability to new scenarios. To improve transferability, transferable sequential recommendation studies~\cite{hou2022towards,10.1145/3543507.3583434,10.1145/3580305.3599519,MoRec} explore leveraging textual features to enhance item representations. These approaches improve the transferability and robustness of recommender systems.
Nowadays, LLMs offer new opportunities for sequential modeling, promising more robust and generalizable recommender systems.

\noindent$\bullet$\quad\textbf{LLM-Based Recommendation.}
%\subsection{LLM-Based Recommendation}
With the rise of LLMs, interest in LLM-based recommender systems has grown, leveraging LLMs as core engines. Early studies~\cite{dai2023uncovering,sanner2023large,wang-etal-2023-rethinking-evaluation} explore their zero-shot/few-shot potential via in-context learning~\cite{dong2024survey}. However, the gap between LLMs' pretraining on general text and recommendation-specific needs leads to suboptimal performance. To address this, recent research follows two paradigms: breadth-oriented and depth-oriented.
The former integrates multi-domain~\cite{10.1145/3705727,peng2024ecellm} or multi-task~\cite{geng2022recommendation,10.1145/3708882,cui2022m6,peng2024ecellm} recommendation data to construct extensive recommendation world knowledge, paving the way for developing a generalizable LLM-based recommender. For example, Peng et al.\cite{peng2024ecellm} build large-scale e-commerce instruction dataset ECInstruct and develop generalist LLM for e-commerce. P5~\cite{geng2022recommendation} designs prompts to unify 5 recommendation tasks and presents a unified text-to-text recommendation paradigm.
Depth-oriented paradigm seeks to enable LLMs to deeply comprehend recommendation tasks within specific domains. 
Key areas of focus include: 1) the in-depth alignment of domain-specific recommendation knowledge, such as collaborative signals~\cite{lin2024bridging,10.1145,10.1145/3626772.3657690,10.1145/3589334.3645458,kong2024customizing}.
The introduction of collaborative signals effectively improves model performance in warm-start scenarios. However, this comes at the cost of reduced generalization, making it challenging to adapt to new domains and cold-start situations.
Another research focus is 2) the development of efficient alignment methods between large language models and recommendation tasks. These methods include leveraging~\cite{bao2023tallrec,kong2024customizing} LoRA fine-tuning techniques and designing data-efficient fine-tuning strategies~\cite{10.1145/3626772.3657807, zheng2024harnessing}.
In fact, the two paradigms have complementary advantages. In this work, we investigate how to integrate the advantages of both paradigms to simultaneously achieve both breadth and depth.

\iffalse

Existing LLM-based recommendation methods can be broadly classified into two categories based on the specific emergent capabilities of LLMs they prioritize. (1) The first category aims to build general-purpose recommender systems by harnessing the generalizability of LLMs~\cite{10.1145/3705727,geng2022recommendation,10.1145/3708882,peng2024ecellm,dai2023uncovering}.
This line of research reformulates data from diverse recommendation tasks or domains into unified textual prompts, which are then used as input for LLMs.
Early works leveraging LLMs for recommender systems, aiming to leverage the LLMs' rich world knowledge, strong reasoning, and generalization abilities.

Existing research on LLMs for recommendation can be primarily divided into two categories: (1) LLM-enhanced recommendations, which treat LLMs as powerful feature extractors to enrich user and item representations. (2) LLM-based recommendations, which directly employ LLMs as recommender systems. Along the second line, early works xxxxxx
\fi

\noindent$\bullet$\quad\textbf{Model Merging.}
%\subsection{Model Merging}
Model merging aims to combine multiple expert models into a more capable single model, offering benefits such as reduced storage and serving costs~\cite{yang2024modelmergingllmsmllms}.
Model merging techniques are applied in various scenarios such as unlearning old-knowledges in LLMs~\cite{zaman-etal-2024-fuse}, understanding content across multiple modalities~\cite{aiello2024jointly,chen-etal-2024-model}, generating images with different styles or achieving image-style transformation~\cite{biggs2024diffusion}.
Previous attempts involve merging multiple models, all initially trained on the same task, with the aim of enhancing the model’s overall generalization~\cite{Gupta2020Stochastic,wang2022meta}. 
Inspired by these works, we apply model merging to the LoRA modules, leveraging their ability to integrate domain-general and domain-specific knowledge effectively.